\documentclass[aps, a4paper, twocolumn, nofootinbib, showpacs, amsfonts]{revtex4}

\usepackage[utf8]{inputenc}
\usepackage{ dsfont }
\usepackage{amssymb}

\usepackage{graphicx}
\usepackage{amsmath}
\usepackage{bm}
\usepackage{xcolor}

\usepackage[toc,title]{appendix}

\usepackage{longtable}
\usepackage{epsfig}
\usepackage{amssymb}
\usepackage{amsmath}
\usepackage{verbatim}
\usepackage{xspace}
\usepackage{titlesec}
\usepackage{mathtools}
\usepackage{multirow}
\usepackage{float}
\usepackage{graphicx}
\usepackage{placeins}
\usepackage{rotating}

\usepackage{tikz}
\usepackage{mathrsfs}
\usetikzlibrary{arrows}

\usetikzlibrary{calc,decorations.pathmorphing,shapes}

\usepackage{fancyhdr}
\usepackage[utf8]{inputenc}
\newcommand{\bea}{\begin{eqnarray}}
\newcommand{\eea}{\end{eqnarray}}
\newcommand{\be}{\begin{equation}}
\newcommand{\ee}{\end{equation}}

\definecolor{rvwvcq}{rgb}{0.08235294117647059,0.396078431372549,0.7529411764705882}
\definecolor{wrwrwr}{rgb}{0.3803921568627451,0.3803921568627451,0.3803921568627451}

\allowdisplaybreaks[1]

\begin{document}

\title{On the ultraviolet behavior of conformally reduced quadratic gravity}.

\author{Alfio Bonanno}
\email{alfio.bonanno@inaf.it}

\affiliation{INAF, Osservatorio Astrofisico di Catania, via S.Sofia 78, I-95123 Catania, Italy}

\author{Maria Conti}
\email{mconti@uninsubria.it}

\affiliation{DISAT, Università degli Studi dell’Insubria, via Valleggio 11,  I-22100 Como, Italy\\INFN, Sezione di Milano, Via Celoria 16, I-20133 Milano, Italy }

\author{Sergio Cacciatori}
\email{sergio.cacciatori@uninsubria.it}

\affiliation{DISAT, Università degli Studi dell’Insubria, via Valleggio 11,  I-22100 Como, Italy\\INFN, Sezione di Milano, Via Celoria 16, I-20133 Milano, Italy }

\begin{abstract}
We study the conformally reduced $R+R^2$ theory of gravity and we show that the theory is asymptotically safe with an ultraviolet critical manifold of dimension three.  In particular, we discuss the universality properties of the fixed point and its stability under the use of different regulators with the help of the proper-time flow equation.  
We find three relevant directions, corresponding to the $\sqrt{g}$, $\sqrt{g} R$ and $\sqrt{g} R^2$ operators, whose 
critical properties are very similar to the ones shared by the full theory. 
Our result shows that the basic mechanism at the core of the Asymptotic Safety program is still well described by the conformal sector also beyond the Einstein-Hilbert truncation.  Possible consequences for the asymptotic safety program are discussed. 
\end{abstract}

\maketitle

\section{Introduction}

Despite the fact that the Asymptotically Safe approach to quantum gravity is a relatively new approach to the quantization of the gravitational field, the basic assumptions of this approach are deeply rooted in old-fashioned, standard Quantum Field Theory. 
The basic idea at the core of this program is that, as long as we do not insist on the notion of continuum limit tailored to perturbation theory, gravity can be treated along the same lines as similar quantum field theories whose continuum limit is defined non-perturbatively.  This possibility was first suggested by Weinberg  \cite{1979grec.conf..790W}  and further implemented by Reuter  \cite{Reuter:1996cp} by means of  the Wilsonian renormalization group (RG) formalism in Quantum Field Theory (QFT)   \cite{Niedermaier:2006wt,Bonanno:2009nj,Litim:2011cp,Nagy:2012ef,Eichhorn:2017egq,Pereira:2019dbn,Reichert:2020mja,Bonanno:2020bil,Platania:2020lqb}.  (See the recent books \cite{Percacci:2017fkn,Reuter:2019byg} for a pedagogical introduction to the subject.)

It is possible to explain the technical mechanism which lies at the root of the non-perturbative renormalization of Einstein's gravity in simple physical terms. Perhaps the most illuminating discussion in this context has been presented by Polyakov \cite{Polyakov:1993tp}, who noticed that as gravity is always attractive and  therefore a  larger cloud of virtual particle implies a stronger gravitational force,  Newton's constant $G$ should be {\it antiscreened} at small distances. The implication of this behaviour suggests that  the dimensionless coupling constant $g(\ell) = G(\ell) /\ell^2$  tends to a finite non-zero limit at small distances
\begin{equation}
\label{as}
\lim_{\ell \rightarrow 0} g(\ell) \propto g^\ast \not = 0
\end{equation}
as $G$ scales  as $\ell^2$  according to its natural dimensions. A theory whose dimensionless coupling constant approaches a non-Gaussian (non-vanishing) fixed point (NGFP) in the short distance limit  as in (\ref{as}) is called {\it asymptotically safe} (at variance with the more familiar case of the asymptotic freedom where $g^\ast=0$.)  

In a series of papers Reuters and collaborators have clarified that conformal factor plays a central role in the emergence of the NGFP in the ultraviolet region and in the determination of the critical properties of the theory \cite{Reuter:2008wj,Reuter:2008qx,Manrique:2010am}. There are two issues in particular which make the emergence of \ref{as} highly non-trivial, the first one is the use of the background field approach, and the second is the pivotal role played by the conformal field instability. In fact, the central idea of conformal field quantization is to employ the background metric (in the sense of the background field method) in constructing the Wilsonian renormalization group equations. On the other hand, as the conformal factor has the wrong kinetic sign in the euclidean theory, a special regulator must be employed to cutoff modes with the ``wrong'' ultraviolet stability properties. As first discussed in \cite{Bonanno:2012dg}, the IR evolution of the renormalization trajectory can be problematic and only an ultraviolet evolution can be consistently defined.  Most probably a new kind of perturbative continuum limit for quantum gravity emerges in the deep UV for the conformally reduced theory, in this case, \cite{Mitchell:2020fjy}. 

The role of the conformal degrees of freedom in determining the presence of the NGFP has been discussed for pure $R^2$ in \cite{Machado:2009ph}. The authors found that the fixed point disappears for pure gravity but is present when  matter is included. In four dimensions the question was further discussion in a conformally reduced version of the $f(R)$ theory in \cite{Demmel:2015oqa} . The authors found a scaling solution with properties qualitatively very similar to the ones constructed from the full flow equation obtained using a smooth approximation to the full spectrum of the hyperspherical harmonics.  More recently the existence of the NGFP for the conformally reduced $R+R^2$ theory has been questioned in \cite{Knorr:2020ckv} a flat-space derivative expansion combined with an exponential parametrisation for the fluctuations of the conformal factor has been used.

In this work, close in spirit to the original  \cite{Lauscher:2002sq} calculation, we consider a conformally reduced $R+R^2$ theory with a linear parametrization for conformal factor fluctuations, and  a compact background projection.  We find that the NGFP is clearly present and, more importantly,  the universal properties of the UV critical manifold of the reduced theory are very similar to those shared by the full quadratic theory.  Our conclusion is that the role of the other degrees of freedom of the metric is mostly relevant for the  actual location of the fixed point in the space of the coupling constants, but it plays no significant role in deciding the universal properties of the theory. 

\section{Functional flow equation for the conformal factor}\label{sectionformalism}

It is instructive to clarify the main steps in the derivation of the flow equation for the conformal factor. In fact despite the fact that formally one deals with a scalar theory, the requirement of background independence plays an important role in the definition of the regulator.

Let $S[\chi]$ be the action for the fundamental field $\chi(x)$ that we write as $\chi(x) = \chi_{B}(x) + f(x)$ where $\chi_B(x)$ is a non-dynamical background field and $f(x)$ a dynamical (fluctuating)  field. In this formalism $\chi$ plays the same role of a microscopic metric $\gamma_{\mu\nu}$ in the full theory.  In the complete framework a background metric $\bar{g}_{\mu\nu}$ is chosen in order to perform the actual calculations, and the fluctuations $h_{\mu\nu}$ are thus ``integrated-out" in momentum shell (in the wilsonian approach).  The background should  be dynamically determined by the requirement that the expectation value of the fluctuation field vanishes, 
$\langle h_{\mu\nu}\rangle\equiv \bar h_{\mu\nu}=0$.  
Any physical length must then be proper with respect to the background metric $\bar{g}_{\mu\nu}$.  In the conformally reduced theory the expectation values $\bar f\equiv \langle f\rangle$ and $\phi \equiv \langle \chi \rangle=\chi_B +\bar f$ are the analogs of $\bar h_{\mu\nu}\equiv \langle h_{\mu\nu} \rangle$ and $g_{\mu\nu} = \langle \gamma_{\mu\nu} \rangle = \bar g_{\mu\nu}+\bar h_{\mu\nu}$ in the full theory.

The central idea of the conformal field quantization is to employ the background metric 
\begin{equation}
\bar{g}_{\mu\nu} = \chi_B^{2\nu}\, \hat{g}_{\mu\nu}
\end{equation}
where $\hat{g}_{\mu\nu}$  is a reference metric which plays no dynamical role but it is instead fixed to perform the actual calculations on the geometry defined by $\bar{g}$.  The momentum scale $k$ is therefore a ``proper"-$\bar{g}_{\mu\nu}$ momentum scale defined from the eigenvalues of the $-\bar\Box$ operator.  Therefore
\begin{equation}
\label{kk}
\bar{k}^2= \chi_B^{-2\, \nu}\, k^2
\end{equation}
in the case of a  constant $\chi_B$.
As discussed in \cite{Reuter:2008wj,Bonanno:2012dg} the difference $\Delta S_{k_B} = S_{k_b+\delta k_B} - S_{k_B}$ can then be evaluated in the infinitesimal momentum shell between $k_B$ and $k_B+\delta k_B$, where $k_B$ is the ``proper" momentum operator built with the background metric $\bar{g}_{\mu\nu}$. A functional flow equation is finally obtained by taking the $\delta k_B \rightarrow 0$ limit and performing a renormalization group improvement of the resulting expression.  
After this step is accomplished, the ``background-independent" flow is obtained expressing all the running ``proper" momenta in terms of the reference energy scale $k$.
Rewriting the (regularized) one-loop contribution in the Schwinger ``proper-time" formalism one finds
\small
\begin{equation}
\label{+}
\partial_t\, {S}_k [f; \chi_B] =  - \frac{1}{2}\,  {\rm Tr} \int_0^{\infty} \frac{ds}{s}
\, \partial_t\, \rho_{k}\, \exp \Big\{ -s \frac{\delta^2 S_k [f; \chi_B]}{\delta {f}^2} \Big\} \, ,
\end{equation}
\normalsize
where $t \equiv \log(k)$ is the RG time and $\rho_k = \rho_k[\chi_B]$.
The important difference between this type of functional ``proper-time" flow equation and the version used in earlier investigations is that the trace in (\ref{+}) is here computed by means of  the representation provided by the spectrum of  $-\bar\Box$, 
\small
\begin{equation}
\overline{{\rm Tr}} [A] \equiv \int d^dx\, \sqrt{\bar{g}}\,\, \langle x | A | x \rangle = \int d^dx\, \sqrt{\hat{g}}\,\chi_B^{d \nu}\, \langle x | A | x \rangle\, .
\end{equation}
\normalsize
%
For actual calculations  we shall use various families of smooth cutoffs $\rho_k\equiv \rho_k^{1,2}(s,n)$ that have been widely used in the literature \cite{Bonanno:2012dg,Bonanno:2004sy,Litim:2010tt} whose explicit expressions are
\small
\begin{align}
&\rho_k^1(s,n)=\frac{\Gamma(n,s\, {\cal Z}\, \hat{k}^2)-\Gamma(n,s\, {\cal Z}\, \Lambda^2)}{\Gamma(n)}\label{regform1}\\
&\rho_k^2(s,n)=\frac{\Gamma(n,s\, n\, {\cal Z}\, \hat{k}^2)-\Gamma(n,s\, n \,{\cal Z}\, \Lambda^2 )}{\Gamma(n)}.\label{regform2}
\end{align}
\normalsize
Here $n$ is an arbitrary real, positive parameter that controls the shape of the $\rho_k^{1,2}(s,n)$ in the interpolating regions, and $\Gamma(\alpha,x)=\int^\infty_x dt \; t^{\alpha-1} e^{-t}$ denotes the incomplete Gamma-function. Furthermore, ${\cal Z}$ is a constant which has to be adjusted: being the kinetic terms of the field of type $a$ of the form $-Z_a\hat{\Box}$, we impose exactly ${\cal Z}=Z_a$. With this prescription, in \eqref{regform1} the eigenvalues of $\hat{\Box}$ are cut off at $\sim \hat{k}^2$, instead of $\sim \hat{k}^2/Z_a$. Similarly, in \eqref{regform2} the cutoff is located at $\sim \hat{k}^2/n$. These two choices represent two so-called `spectral adjustments'. Moreover, as background independence must be achieved, the trace inside the flow equation \eqref{+} must be performed on the modes of the background $\bar{g}_{\mu\nu}$. This is concretely performed inside the regularizators through the identification $\hat{k}^2=\chi_B^{2\nu}\bar{k}^2$. Finally, $\Lambda$ represents the cutoff in the UV. As we are interested only in the Wilson-Kadanoff portion of the RG, the UV cut-off is sent to infinity. Overall, this leads to implementing the scaling laws
\small
\begin{align}
&\partial_t \rho^1_k(s,n)=-\frac{2}{\Gamma(n)}(s\, Z\, k^2 \,\chi_B^{2\nu})^n e^{-s\, Z k^2 \chi_B^{2\nu}}\label{1dtreg}.\\
&\partial_t \rho^2_k(s,n)=-\frac{2}{\Gamma(n)}(s\, n\, Z k^2\, \chi_B^{2\nu})^n e^{-s\, n Z k^2 \chi_B^{2\nu}}\label{2dtreg}
\end{align}
\normalsize
inside the flow equation. Concretely, the calculations for both cutoff families are performed through a range of values for the smoothness parameter $n$: ${n=\left\lbrace 3,5,7,9,10,15,20,30,40,50\right\rbrace} $. The limiting case ${n\to\infty}$ is also considered for the second regularizator: this is readily done through the already known identity \cite{Bonanno:2004pq}
\small
\begin{equation}
\lim_{n\to\infty}\partial_t\rho^{(2)}_k(s,n)=-\frac{2}{Z\,k^2\,\chi_B^{2\nu}}\delta\left(s-\frac{1}{Z\,k^2\,\chi_B^{2\nu}}\right).\label{identinfty}
\end{equation}
\normalsize
Finally, as an extra check for the robustness of the physical results under a change in the regularizator shapes, a slight modification in the cutoff structure $\rho_k^{1,2}$ is also studied. Cutoffs \eqref{1dtreg} and \eqref{2dtreg} are built through the regularization on $\bar{k}^2$, which are the modes of the $-\bar{\Box}$ operator. Because our model (described in detail in Section \ref{sectioncalcoli}) contains both $-\bar{\Box}$ and $\bar{\Box}^2$ operators, following the reasonings of \cite{Buccio:2022egr}, we also apply a regularization on the quadratic operator in order to check possible differences in the physics of the results. New cutoffs are defined:
\small
\begin{align}
&\tilde{\rho}_k^1(s,n)=\frac{\Gamma(n,s\, (Z \hat{k}^2+Z^2 \hat{k}^4))-\Gamma(n,s\,(Z \Lambda^2+Z^2 \Lambda^4))}{\Gamma(n)}\label{regnewform1}\\
&\tilde{\rho}_k^2(s,n)=\frac{\Gamma(n,s\, n\, (Z \hat{k}^2+Z^2 \hat{k}^4))-\Gamma(n,s\, n\, (Z \Lambda^2+Z^2 \Lambda^4))}{\Gamma(n)},\label{regnewform2}
\end{align}
\normalsize
Requiring background independence by cutting the modes of the background metric and sending the UV cutoff $\Lambda\to\infty$ leads to the the following scaling laws:
\small
\begin{align}
&\partial_t \tilde{\rho}^1_k(s,n)=-\frac{2}{\Gamma(n)}(s\, (Z k^2\chi_B^{2\nu}+Z^2 k^4\chi_B^{4\nu}))^n\times\nonumber\\
&\quad\quad\quad\quad\quad\quad\times e^{-s\,(Z k^2\chi_B^{2\nu}+Z^2 k^4 \chi_B^{4\nu})}\label{1dtregnew}.\\
&\partial_t \tilde{\rho}^2_k(s,n)=-\frac{2}{\Gamma(n)}(s\, n\, (Z k^2\chi_B^{2\nu}+Z^2 k^4\chi_B^{4\nu}))^n\times\nonumber\\
&\quad\quad\quad\quad\quad\quad\times e^{-s\, n\, (Z k^2\chi_B^{2\nu}+Z^2 k^4\chi_B^{4\nu})}\label{2dtregnew}.
\end{align}
\normalsize

\section{Non-perturbative $\beta$-functions}\label{sectioncalcoli}

In this section implement the RG flow equation approach to study the conformal sector of the the following theory:
\begin{equation}
S=\int d^d x \sqrt{g}\left[\frac{1}{16\pi G}(-R+2\Lambda)+\beta R^2\right].\label{CREHgen}
\end{equation}
A Weyl rescaling $g_{\mu\nu}=\phi(x)^{2\nu}\hat{g}_{\mu\nu}$ is implemented, where $\nu=2/(d-2)$ and $\hat{g}_{\mu\nu}$ is a reference metric. Weyl rescaling leads to
\small
\begin{align}
R=\phi^{-2\nu}\left(\hat{R}-\frac{2\nu(d-1)\hat{\Box}\phi}{\phi}+g(d)\frac{\hat{g}^{\mu\nu}\partial_\mu \phi \partial_\nu\phi}{\phi^2}   \right),\label{Rd}
\end{align}
\normalsize
where ${g(d)=\left(2\nu(d-1)-\nu^2(d-1)(d-2)\right)=0}$ as ${\nu=2/(d-2)}$ and ${\hat{R}=R(\hat{g})}$. Being the scaling of the metric $\sqrt{g}=\phi^{d\nu}\sqrt{\hat{g}}$, the effective action is
\small
\begin{align}
\Gamma_k=\int d^dx \sqrt{\hat{g}}&\left[-\frac{1}{2}Z_k\phi\hat{\Box}\phi+U[\phi]+\right.\nonumber\\
&+16\left(\frac{d-1}{d-2}\right)^2\beta_k(\hat{\Box}\phi)^2\phi^{\frac{2d-8}{d-2}-2}+\nonumber\\
&\left.-8\left(\frac{d-1}{d-2}\right)\beta_k\hat{R}(\hat{\Box}\phi)\phi^{\frac{2d-8}{d-2}-1}\right],\label{Gammakd}
\end{align}
\normalsize
where ${U[\phi]=AZ_k\hat{R}\phi^2-2AZ_k\Lambda_k \phi^{\frac{2d}{d-2}}+\beta_k\hat{R}^2\phi^{\frac{2d-8}{d-2}}}$, $A=(d-2)/8(d-1)$ and \small $$Z_k=-\frac{1}{2\pi G_k}\frac{d-1}{d-2}.$$ \normalsize The expansion of the conformal factor $\phi=\chi_B+\bar{f}$ with $\bar{f}=0$ leads to the l.h.s. of the flow equation:
\small
\begin{align}
\partial_t\Gamma_k=\int d^d x \sqrt{\hat{g}}&\left[(\partial_t Z_k)A \hat{R}\chi_B^2-2A\partial_t(Z_k\Lambda_k)\chi_B^{\frac{2d}{d-2}}+\right.\nonumber\\
&+\left.(\partial_t\beta_k)\hat{R}^2\chi_B^{\frac{2d-8}{d-2}}\right].\label{lhsd}
\end{align}
\normalsize
The explicit expression for the r.h.s. requires the computation of the contribution of the second functional derivative of the effective action inside the trace. From the expression of effective action \eqref{Gammakd} it is possible to write its second functional derivative. Because the field is then expanded as $\phi=\chi_B+\bar{f}$ with constant background $\chi_B$ and fluctuation $\bar{f}\to 0$, we can write $\Gamma_k^{(2)}$ as the sum of two terms:
\begin{equation}
\Gamma_k^{(2)}=\mathcal{X}+\mathcal{Y},
\end{equation}
where $\mathcal{X}$ indicates the terms surviving $\bar{f}\to 0$, while the operator $\mathcal{Y}$ will contain only terms involving powers of $\bar{f}$ or derivatives of $\bar{f}$, which will be null when considering $\bar{f}\to 0$, hence adding no contribution to the present calculation. We are only left with the $\mathcal{X}$ operator, which can be written as:
\small
\begin{align}
\mathcal{X}&=2Z_k A\hat{R}-2ABZ_K\Lambda_k\chi_B^{\frac{2d}{d-2}-2}+C\beta_k\hat{R}^2 \chi_B^{\frac{2d-8}{d-2}-2}\nonumber+\\
&-\left(Z_k+16\left(\frac{d-1}{d-2}\right)\left(\frac{2d-8}{d-2}-1\right)\beta_k\hat{R}\chi_B^{\frac{2d-8}{d-2}-2}\right)\hat{\Box}\,+\nonumber\\
&+32\left(\frac{d-1}{d-2}\right)^2\beta_k \chi_B^{\frac{2d-8}{d-2}-2}\hat{\Box}^2.\label{Gammak2d}
\end{align}
\normalsize
This relatively short expression for the second functional derivative of the effective action is due to the choice of the projection onto a spherical geometry, where the $\mathcal{Y}$ operator can be dismissed right away. In a flat geometry, for example, $\hat{R}=0$ and the effective action \eqref{Gammakd} would look very different. In this case, the terms involving derivatives of the fluctuation field should be carried on throughout the calculation in order to find their respective equivalents in the r.h.s. of the flow equation: this would require the computation of $\mathcal{Y}$ and its evaluation inside the trace - a lengthy task, as $\mathcal{Y}$ will contain many non-commuting operators, especially going at higher orders in curvature. The choice of the spherical geometry corresponds instead with a constant, non-null curvature, which allows one to readily forget about all the operators involving derivatives of the fluctuation field, which can be considered null at earlier steps in the computation. The original pure CREH study \cite{Bonanno:2012dg} considered two different geometry projections ($\mathbb{S}^d$ spherical and $\mathbb{R}^d$ flat geometry) in order to perform comparisons on the results and identify `universal quantities', i.e. quantities unaltered by external factors such as the choice of the geometry projection: following the same spirit, we consider the spherical projection but implement different regularization schemes in the cutoff of the IR spectrum.
In particular, we implement the scaling laws \eqref{1dtreg}-\eqref{2dtreg} of two different regularization families \eqref{regform1}-\eqref{regform2} already introduced in Section \ref{sectionformalism}. As the trace must be performed on the modes of the operator $-\bar{\Box}$, each $-\hat{\Box}$ in the operator $\mathcal{X}$ and each mode $\hat{k}$ in the regularizators must be expressed in terms of their background counterparts through $\hat{\Box}=\chi_B^{2\nu}\bar{\Box}$ and $\hat{k}^2=\chi_B^{2\nu}\bar{k}^2$. For example, using the second regularizator $\rho_k^2(s,n)$ leads to the following r.h.s. of the flow equation:
\small
\begin{align}
&\partial_t\Gamma_k=\frac{1}{2}\int_0^\infty\frac{ds}{s}\left(-\frac{2}{\Gamma(n)}(s\, n\, Z_k \chi_B^\frac{4}{d-2} k^2)^n e^{-s\, n\, Z_k \chi_B^{\frac{4}{d-2}}k^2}\right)\times\nonumber\\
&\times e^{-s\left[Z_k\left(2A\hat{R}-2AB\Lambda_k \chi_B^{\frac{2d}{d-2}-2}\right)+C\beta_k\hat{R}^2 \chi_B^{\frac{2d-8}{d-2}-2}\right]}\overline{\text{Tr}}W(-\bar{\Box}),\label{rhsprogression}
\end{align}
\normalsize
where
\small
\begin{align}
&W(-\bar{\Box})=e^{-s\left[-\left(Z_k\chi_B^{\frac{4}{d-2}}+D\beta_k\hat{R}\chi_B^{\frac{2d-4}{d-2}-2}\right)\bar{\Box}+E\beta_k \chi_B^{\frac{2d}{d-2}-2}\bar{\Box}^2\right]},
\end{align}
with $D=16\left(\frac{d-1}{d-2}\right)\left(\frac{2d-8}{d-2}-1\right)$ and $E=32\left(\frac{d-1}{d-2}\right)^2$.
\normalsize
The last trace can be performed through a Seeley-Gilkey-deWitt heat kernel expansion at quadratic order in curvature:
\small
\begin{align}
\overline{\text{Tr}}[W(-\bar{\Box})]&=\frac{1}{(4\pi)^{d/2}}\int d^d x \sqrt{\bar{g}}\sum_{k \geq 0}[a_k]Q_{\frac{d}{2}-k},\label{formulanostra}
\end{align}
\normalsize
where $Q_{n}=\frac{1}{\Gamma[n]}\int_0^\infty dz W(z)z^{n-1}$ and where the first coefficients of the expansion in the spherical projection are $[a_0]=1$, $[a_1]=\hat{R}/{6}$, $[a_2]=f(d)\hat{R}^2$ with
\small
\begin{equation}
f(d)=\frac{1}{18}\left(\frac{1}{4}+\frac{1}{5d(d-1)}-\frac{1}{10 d}\right).
\end{equation}
\normalsize
With this expansion, the r.h.s. of the flow equation can be written as the sum of three integrals: for example, in the case of the second regularizator it becomes explicitly 
\small
\begin{equation}
\partial_t\Gamma_k=I_1+\frac{\hat{R}}{6}I_2+\hat{R}^2f(d)I_3\label{rhsshort}
\end{equation}
\normalsize
where each integral $I_j$ (with $j=1,2,3$) is defined as
\small
\begin{align}
I_j=\int d^d x& \sqrt{\hat{g}}\,\chi_B^{d\nu}\frac{1}{(4\pi)^{d/2}}(Z_k n k^2 \chi_B^{\frac{4}{d-2}})^n\times\nonumber\\
&\times\frac{1}{\Gamma(\frac{d}{2}+1-j)}\int_0^\infty dq\,\frac{q^{\frac{d}{2}-j}}{\left[a+bq+cq^2\right]^n},\label{casinod}
\end{align}
\normalsize
where
\small
\begin{align}
&a=Z_k n k^2 \chi_B^{\frac{4}{d-2}}+Z_k(2A\hat{R}-2AB\Lambda_k\chi_B^{\frac{2d}{d-2}-2})+\nonumber\\
&\quad+C\beta_k\hat{R}^2\chi_B^{\frac{2d-8}{d-2}-2},\nonumber\\
&b=Z_k\chi_B^{\frac{4}{d-2}}+D\beta_k\hat{R}\chi_B^{\frac{2d-4}{d-2}-2},\nonumber\\
&c=E\beta_k \chi_B^{\frac{2d}{d-2}-2}.\nonumber
\end{align}
\normalsize
Notice that for $d=4$ and for $j=3$, the integral in the $q$ variable in \eqref{casinod} is seemingly divergent in $q=0$. This causes no problem, as this apparent divergence is cured by the opposite divergence given by the factor $1/\Gamma(3-j)$. After having performed the $I_j$ integrals, the r.h.s. \eqref{rhsshort} is then expanded in powers of $\chi_B$. The trajectories $\partial_t Z_k$, $\partial_t(Z_k\Lambda_k)$ and $\partial_t\beta_k$ in \eqref{lhsd} are then recovered through the identification with the corrensponding terms $\hat{R}\chi_B^2$, $\chi_B^{{2d}/(d-2)}$ and $\hat{R}^2\chi_B^{(2d-8)/(d-2)}$ respectively. Finally, the beta functions for the dimensionless couplings $z_k$, $\lambda_k$ and $\beta_k$ can be recovered, where $Z=k^{d-2}\,z$, $\Lambda=k^2\,\lambda$ and $\beta=k^{d-4}\,b$.\\
\\
Because of the exceeding length in computation time for general dimension $d$ and regularizator smoothing parameter $n$, the computation was performed fixing the value $d=4$ and different finite values of $n$: ${n=\left\lbrace3,5,7,9,10,15,20,30,40,50\right\rbrace\,}$, as already mentioned in Section \ref{sectionformalism}. This was done for both regularizators $\rho_k^{1,2}$, where we expect no significant differences in the physics described by the two different cutoffs. Some examples of resulting beta functions can be found in Appendix \ref{APPbetafuncs}. While the procedure outlined in this section is related to finite values of smoothness parameter $n$, the limiting case $n\to\infty$ was also studied for the second cutoff family. This is readily performed through the identity \eqref{identinfty}. With this scaling law, the computations of the r.h.s. of the flow equation are straightforward. The r.h.s. is then
\small
\begin{align}
&\partial_t\Gamma_k=\int d^d x  \sqrt{\hat{g}}\chi_B^{d\nu}\frac{2\pi^{d/2}}{(2\pi)^d \Gamma(d/2)}\int_0^\infty\frac{ds}{s}\frac{1}{Z_k k^2 \chi_B^{\frac{4}{d-2}}}\nonumber\\
&\times\delta\left(s-\frac{1}{Z_k k^2\chi_B^{\frac{4}{d-2}}}\right)\times\nonumber\\
&\times e^{-s \left(2Z_k A\hat{R}-2ABZ_k\Lambda_k\chi_B^{\frac{2d}{d-2}-2}+C\beta_k\hat{R}^2 \chi_B^{\frac{2d-8}{d-2}-2}\right)}\text{Int}_p(s),\label{rhsinftyd}
\end{align}
\normalsize
where $\text{Int}_p(s)$ is defined as the integral
\small
\begin{equation}
\int_0^\infty dp\,p^{d-1} e^{-s\left[\left(Z_k\chi_B^{\frac{4}{d-2}}+D\beta_k\hat{R}\chi_B^{\frac{2d-4}{d-2}-2}\right)p^2+E\beta_k \chi_B^{\frac{2d}{d-2}-2}p^4\right]}.
\end{equation}
\normalsize
The result of the computation of \eqref{rhsinftyd} is then expanded in powers of $\hat{R}$ up to quadratic order and in powers of the background field $\chi_B$ to identify matchings between the r.h.s. and the l.h.s. of the flow equation \eqref{lhsd} as in the finite $n$ case. Examples of the resulting beta functions for the dimensionless couplings related to the $d=4$ case can be found in Appendix \ref{APPbetafuncs}.\\
\\
Finally, the same calculations were performed with the slightly altered cutoff families defined in \eqref{regnewform1}-\eqref{regnewform2}, with corrensponding scaling laws \eqref{1dtregnew}-\eqref{2dtregnew}. As already mentioned in Section \ref{sectionformalism}, with these new scaling laws also the $\bar{\Box}^2$ operator appearing in our model is regularized. The subsequent calculations are extremely similar to the ones shown until now. It is found that the resulting beta functions at fixed dimension $d$ and steepness $n$ perfectly coincide with the beta functions obtained with the original regularizators scaling laws \eqref{1dtreg}-\eqref{2dtreg} at the same $d$ and $n$, highlighting the robustness of the results under a change in the cutoff shape.

\section{Results}\label{Results}
\subsection{Check: CREH limit case $\beta\to 0$}\label{sectioncheck}
As a first check it was shown that, in the limit $\beta\to 0$, the newfound beta functions  reproduce the old beta functions of the pure CREH model studied originally in \cite{Bonanno:2012dg}, together with their UV  attractive fixed points. We remind that critical exponents are defined as the eigenvalues of the stability matrix of the derivatives of beta functions with respect to the different dimensionless couplings, evaluated at the fixed point coordinates:
\[
M_{stab}(\text{FP})=
  \begin{bmatrix}
    \frac{\partial\beta_z}{\partial_z} &  \frac{\partial\beta_z}{\partial_\lambda} \\
    \frac{\partial\beta_\lambda}{\partial_z} &  \frac{\partial\beta_\lambda}{\partial_\lambda}
  \end{bmatrix}\Bigg|_{(z,\lambda)=(z^*,\,\lambda^*)}.
\]
As shown in \cite{Bonanno:2012dg} and Table \ref{tablevaluesold12} (which can be found in Appendix \ref{APPtables}), the pure CREH critical exponents always assume a form of the type $\theta ' \pm i \theta''$, where the Lyapunov exponent $\theta'$ is always a negative number, while the imaginary part $\theta''$ represents a spiral behaviour of the trajectories around the fixed point. The contents of Table \ref{tablevaluesold12} contain the results pertaining the employment of both cutoff families $\rho_k^{1,2}(s,n)$.
As this serves only as a check, a small selection of the finite $n$ values was taken into account: $n=\left\lbrace 3,5,7,9\right\rbrace$. The limit $n\to+\infty$ was also considered for the second cutoff  $\rho_k^2(s,n)$, exploiting the identity \eqref{identinfty} as shown in the previous Section. As expected, the parameters obtained with $\rho_k^2(s,n)$ coincide with the results in \cite{Bonanno:2012dg}, where only the second type of cutoff $\rho_k^2(s,n)$ was studied. Switching between the two different regularizators, it is possible to see that while the coordinates of the fixed points are not preserved by the regularizator switch, the product $g^*\lambda^*$ and the values of the critical exponents represent more universal properties of the CREH model. This was already shown in the original paper \cite{Bonanno:2012dg}, where a comparison between the parameters obtained by the spherical and flat geometry projection was performed. Here the comparison is performed within the same spherical geometry projection, but with different cutoff shapes: this different approach leads nonetheless to the same conclusions of \cite{Bonanno:2012dg}.
\subsection{Results for the conformally reduced quadratic theory}
The beta function systems obtained in the conformal $R+R^2$ theory (some examples of beta functions appear in Appendix \ref{APPbetafuncs}) lead to the determination of a new UV attractive fixed point. This happens in particular for the beta functions related to the choice of $\beta<0$. The properties of this new UV fixed point (such as the coordinates $(g^* (\text{or}\, z^*),\lambda^*,b^*)$, the products of the previous coordinates and the critical exponents $\lambda_1$,$\,\lambda_{2,3}=\theta'\pm i \theta''$) are summarised inside Tables \ref{tablevaluesnew12PT1}-\ref{tablevaluesnew12PT2}) in Appendix \ref{APPtables}.
The new fixed point is characterised by three critical exponents: a real, negative $\lambda_1$ and two complex $\lambda_{2,3}=\theta'\pm i \theta''$, with negative Lyapunov exponent $\theta'$, closely following the CREH behaviour outlined in Section \ref{sectioncheck}. Similarly to the CREH study results, the fixed point coordinates $g^*$ and $\lambda^*$ values do not represent universal quantities per se, as they vary switching between the two different cutoffs $\rho_k^1(s,n)$ and $\rho_k^2(s,n)$ - even though the signs of the coordinates $g^*$ (or $z^*$), $\lambda^*$ and $b^*$ are kept untouched. On the other hand, the coordinate products $g^*\lambda^*$ and $b^*g^*\lambda^*$, the coordinate $b^*$ and the critical exponents do not change with the cutoff choice. The spiral-like behaviour which already appeared in the pure CREH model is still preserved, as shown by the structure of the eigenvalues $\lambda_{2,3}=\theta'\pm i \theta''$.
Note that the signs of the newfound fixed point $g^*<0$ , $\lambda^*>0$ and $b^*<0$ simply correspond to an overall negative sign on the action \eqref{CREHgen} at the fixed point.  Clearly the precise value of the action at the fixed point has no importance in this discussion as we are only interested in discussing the structure of the UV critical manifold, in particular the possibility of defining a non-trivial continuum limit for the conformal factor in the quadratic sector of the full theory. 

\begin{center}
\begin{figure*}
\includegraphics[width = 5. cm]{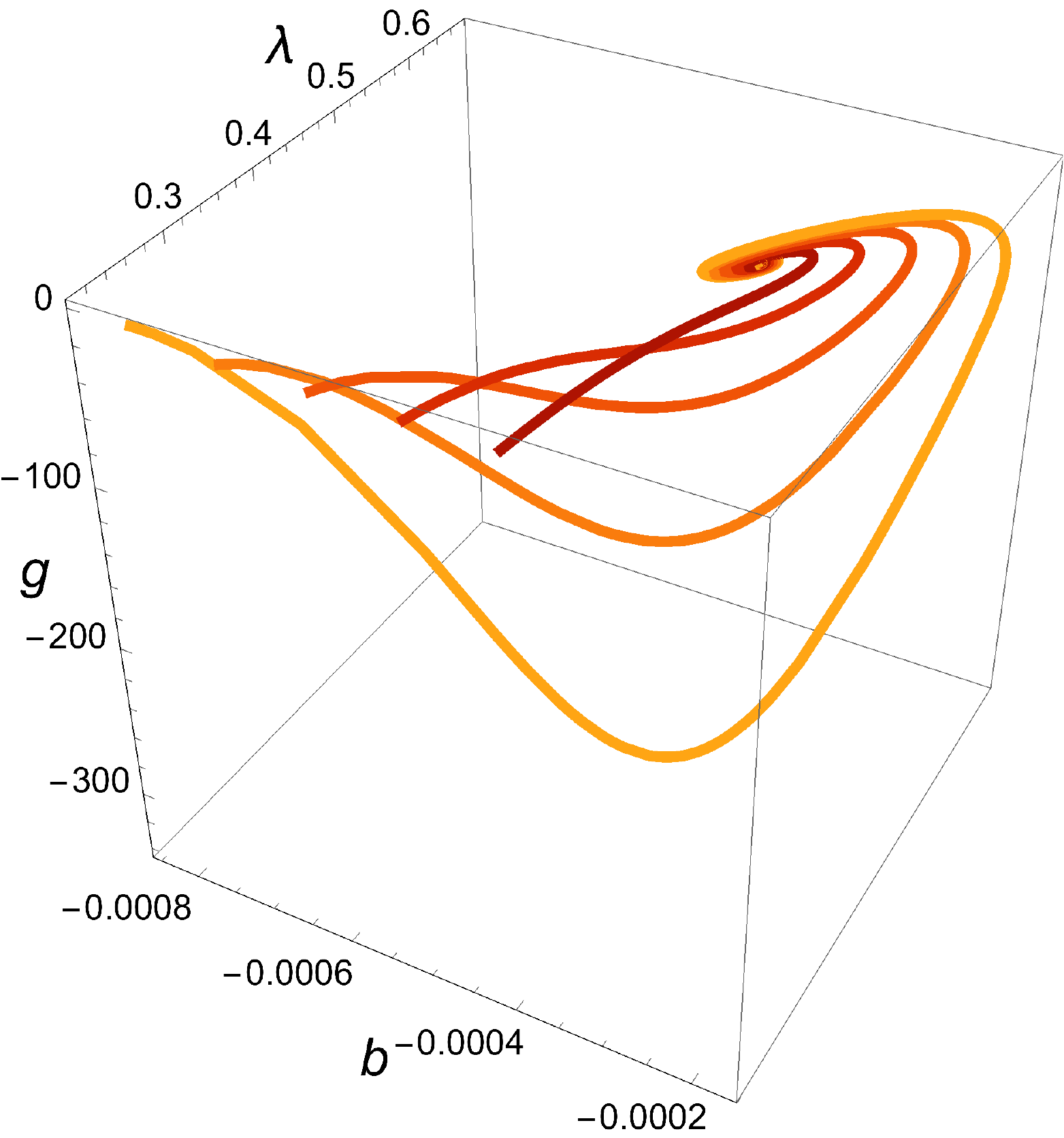}
\includegraphics[width = 5. cm]{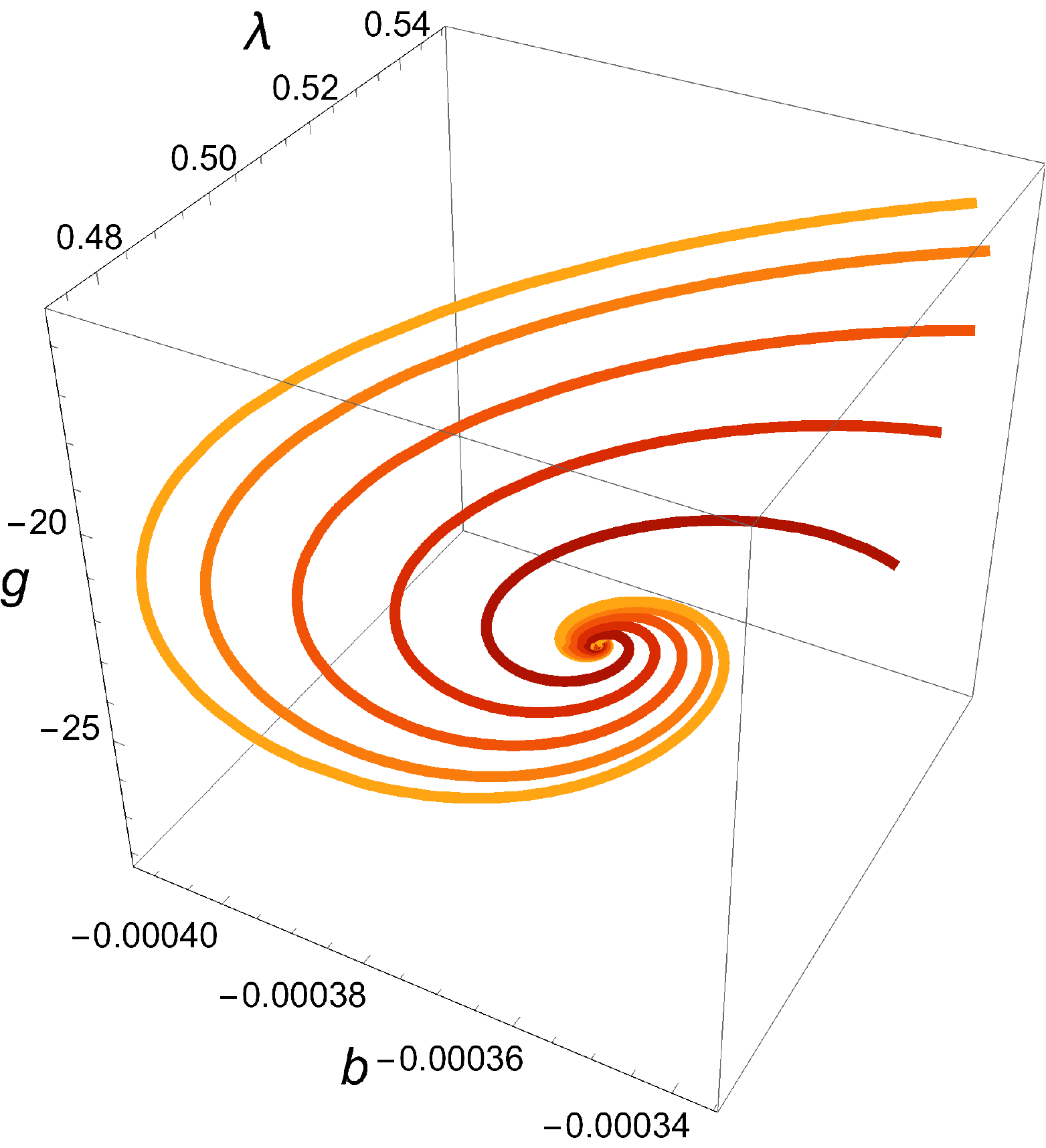}
\includegraphics[width = 5. cm]{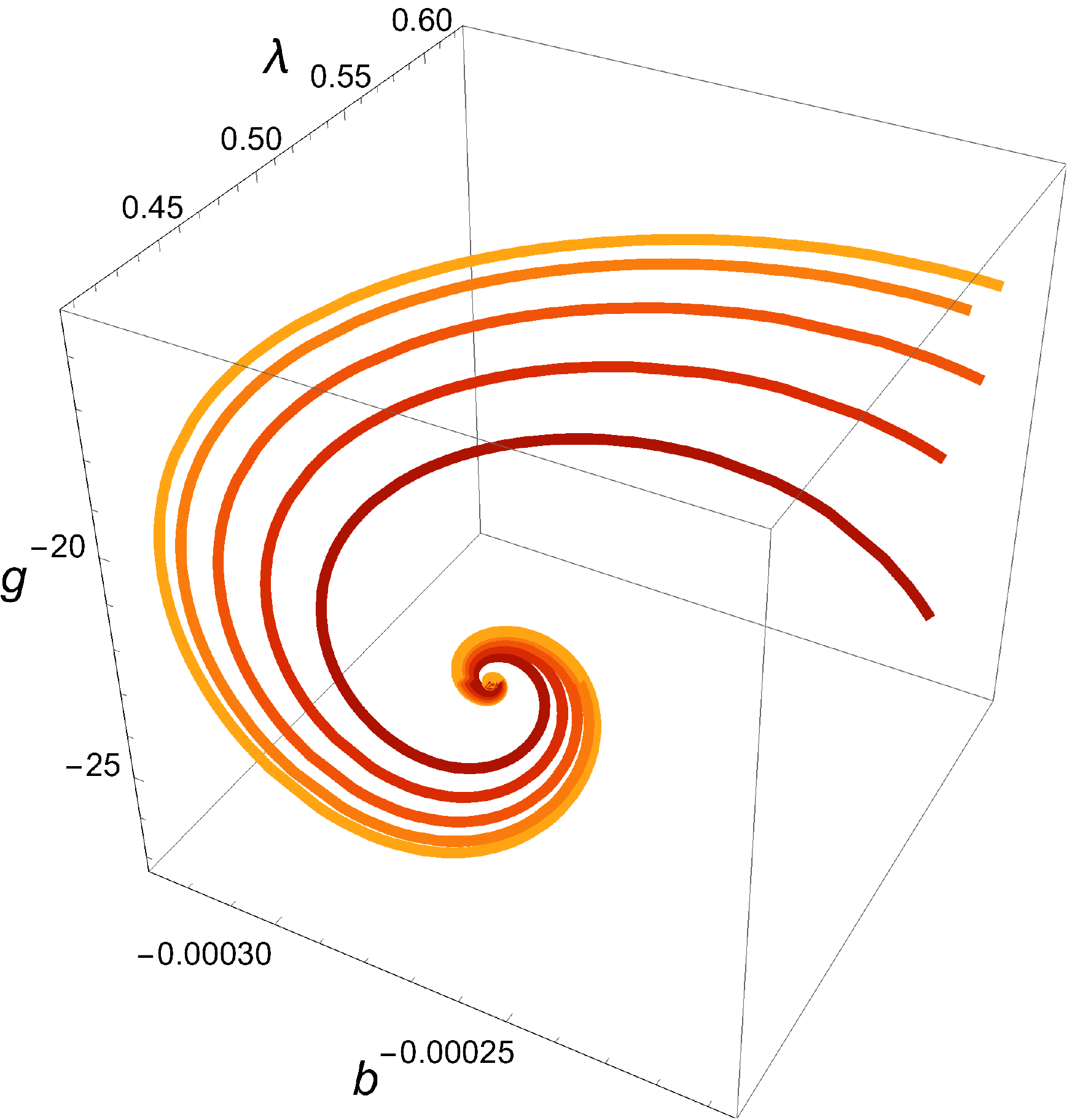}
\caption{Flow of $g$, $\lambda$ and $b$ for various values of the cutoff.}
\label{fig:cycles1}
\end{figure*}
\end{center}

We show in Figure \ref{fig:cycles1} some examples of trajectories in the 3D parameter space obtained by integrating their correspondent beta functions. Each figure depicts the attractive, spiral-like behaviour of the system around their fixed point. In particular, the figure on the left shows how the spiral behaviour flattens onto a plane when it is close to the fixed point: this is represented mathematically by the purely real, negative critical exponent $\lambda_1$ which contributes to the attractiveness of the fixed point, while $\lambda_{2,3}=\theta'\pm i \theta''$ describe the attractive spiral motion component onto a 2D plane.

\begin{figure}[!htbp]
\begin{center}
\includegraphics[scale=0.45]{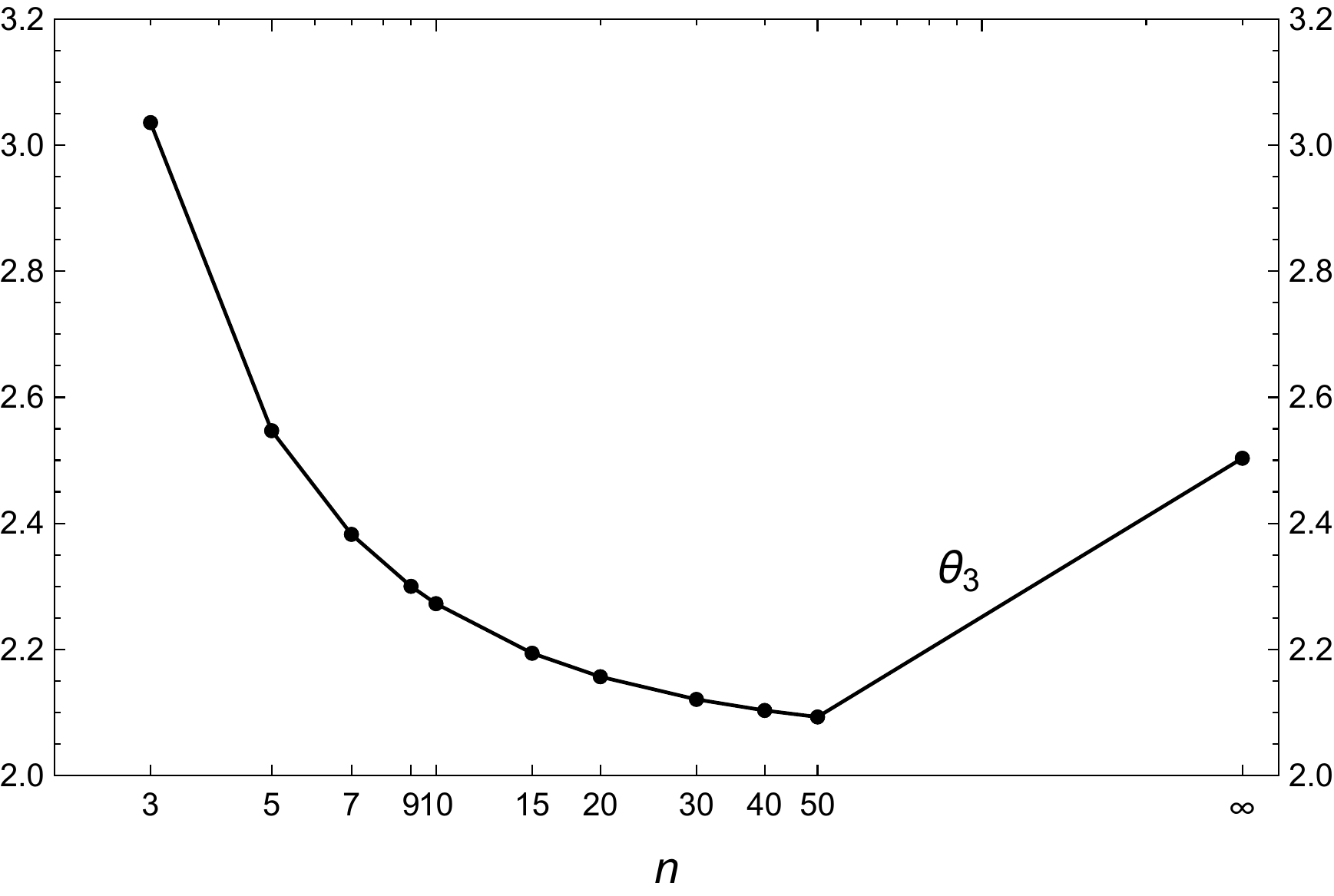}\hfill
\includegraphics[scale=0.45]{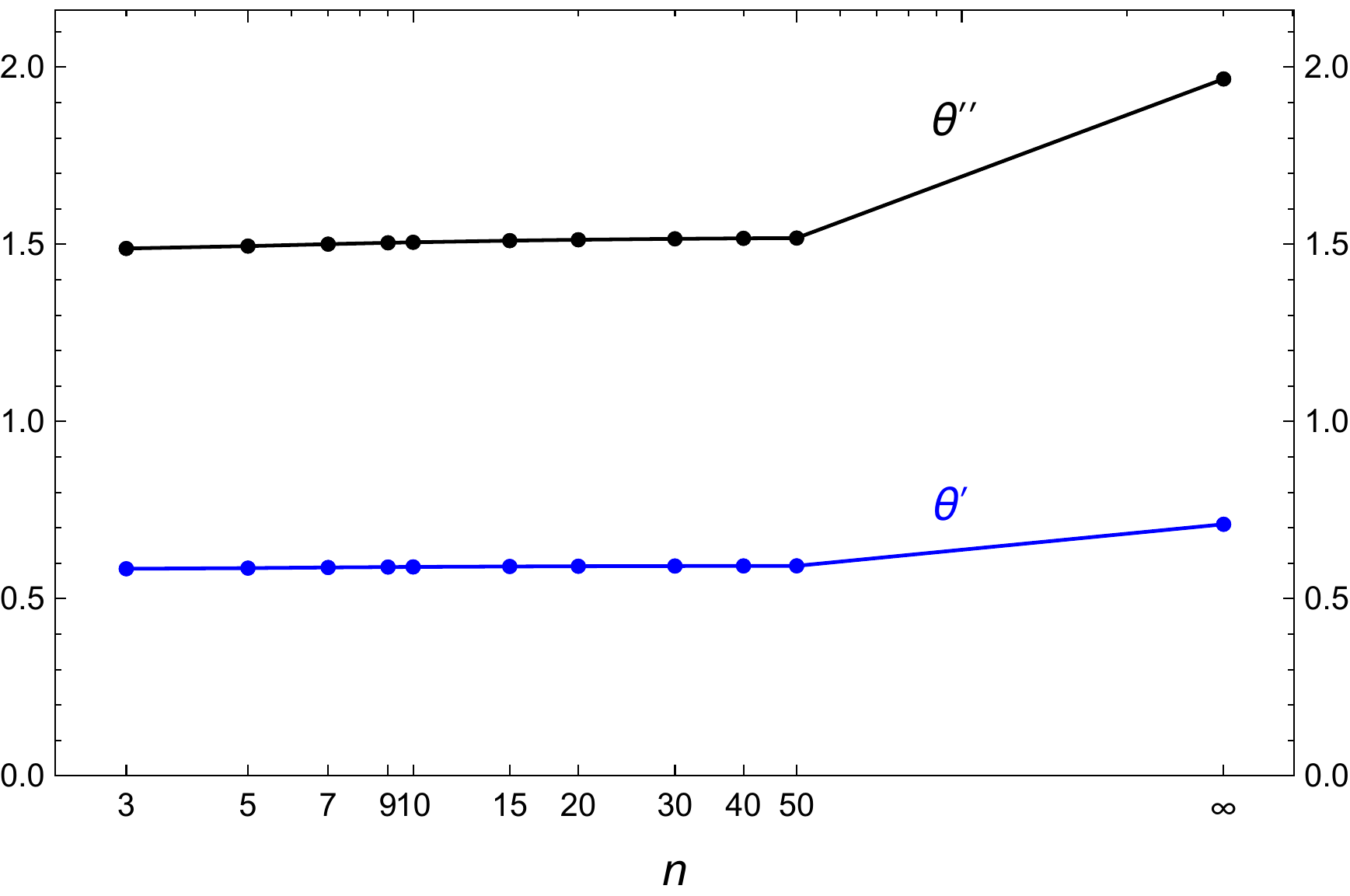}\hfill
\caption{Plot of the critical exponents $\lambda_1$, $\theta'$ and $\theta''$ for different values of $n$. Second cutoff family $\rho^2_k(s,n)$ implemented.}
\label{fig:crit_2reg}
\end{center}
\end{figure}

\begin{figure}[!htbp]
\begin{center}
\includegraphics[scale=0.45]{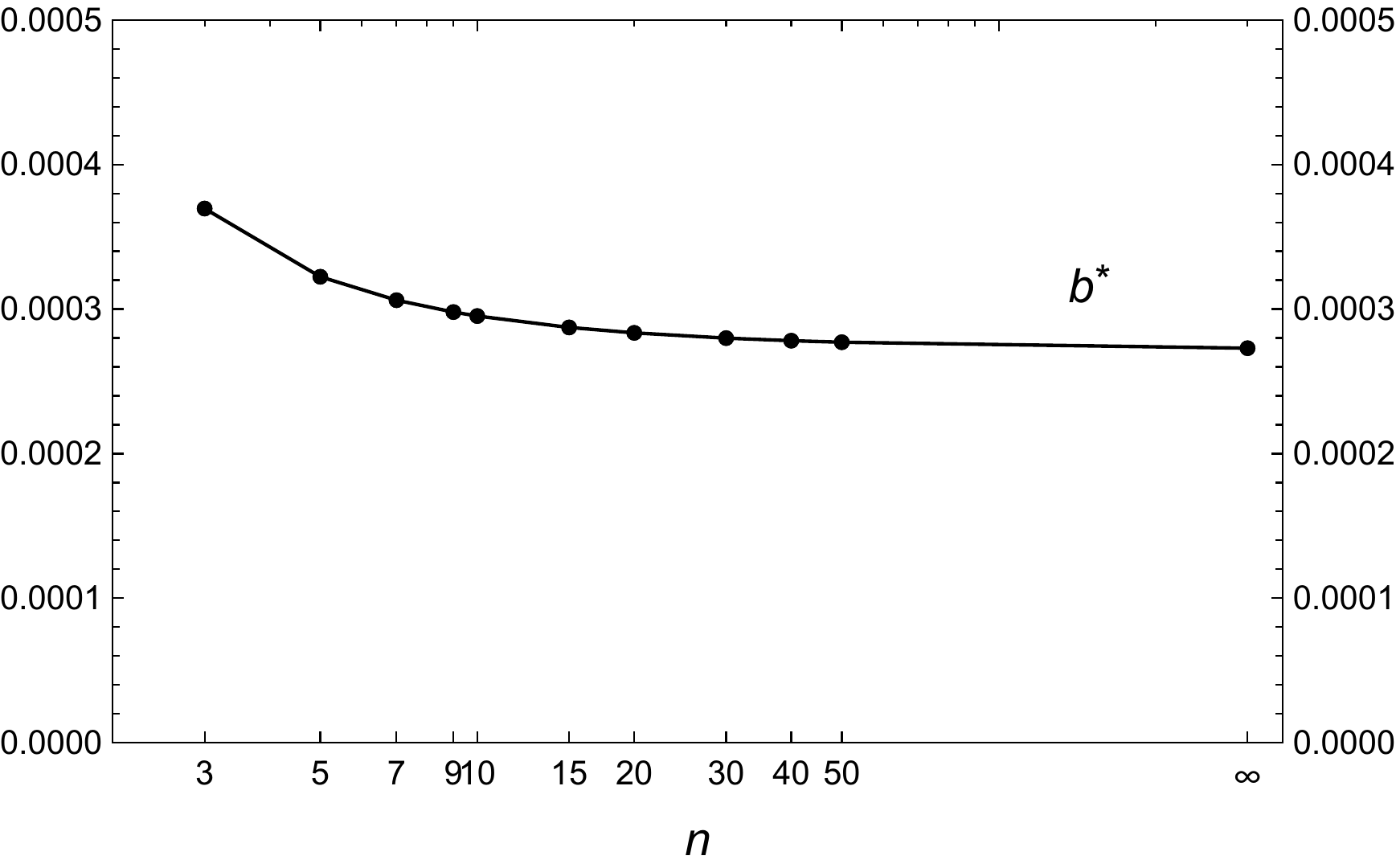}\hfill
\includegraphics[scale=0.45]{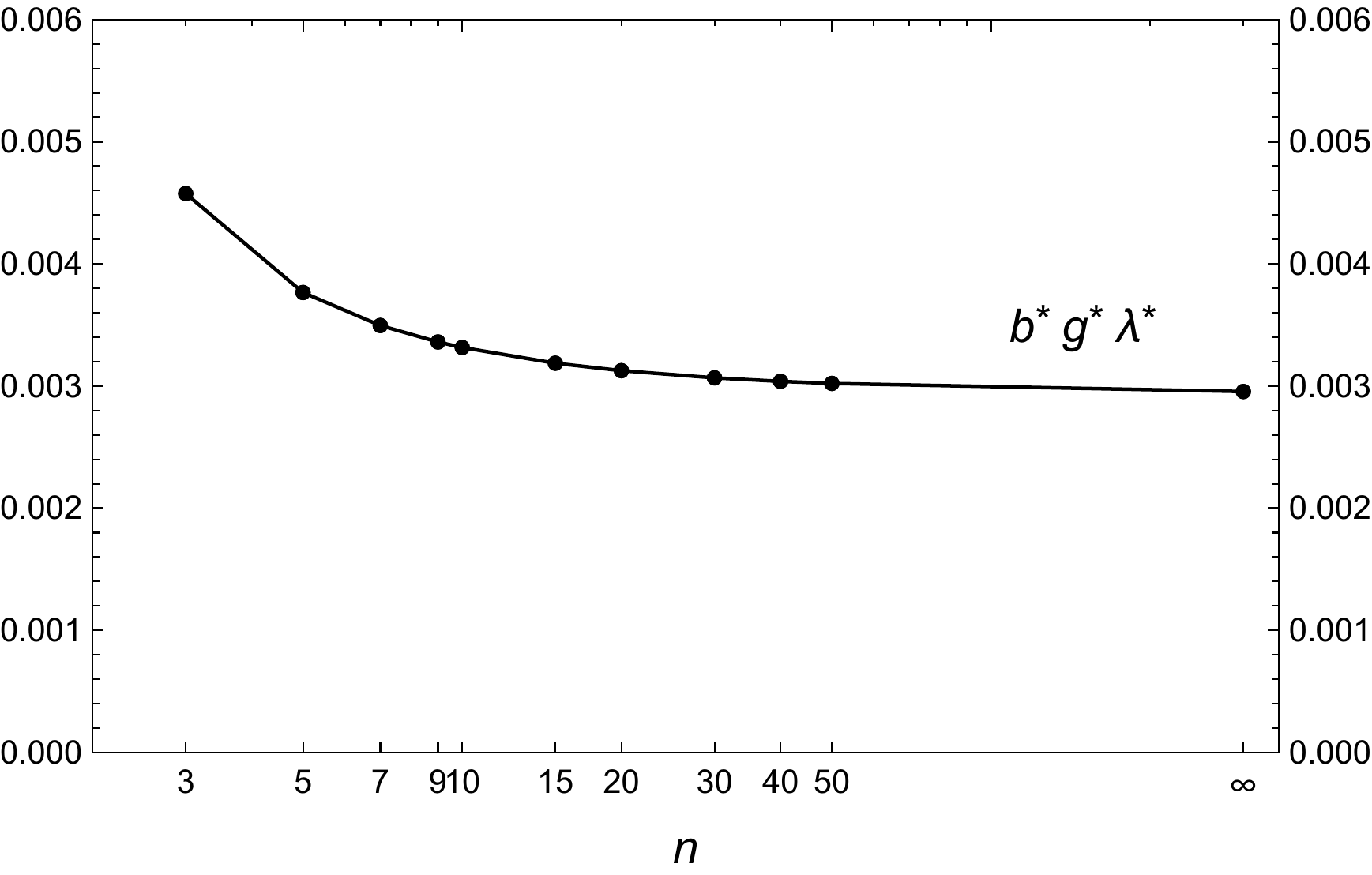}\hfill
\caption{Plot of the fixed point coordinate $b^*$ and the product $b^*g^*\lambda^*$ for different values of $n$. Second cutoff family $\rho^2_k(s,n)$ implemented.}
\label{fig:bandbgl_2reg}
\end{center}
\end{figure}

\begin{figure}[!htbp]
\begin{center}
\includegraphics[scale=0.45]{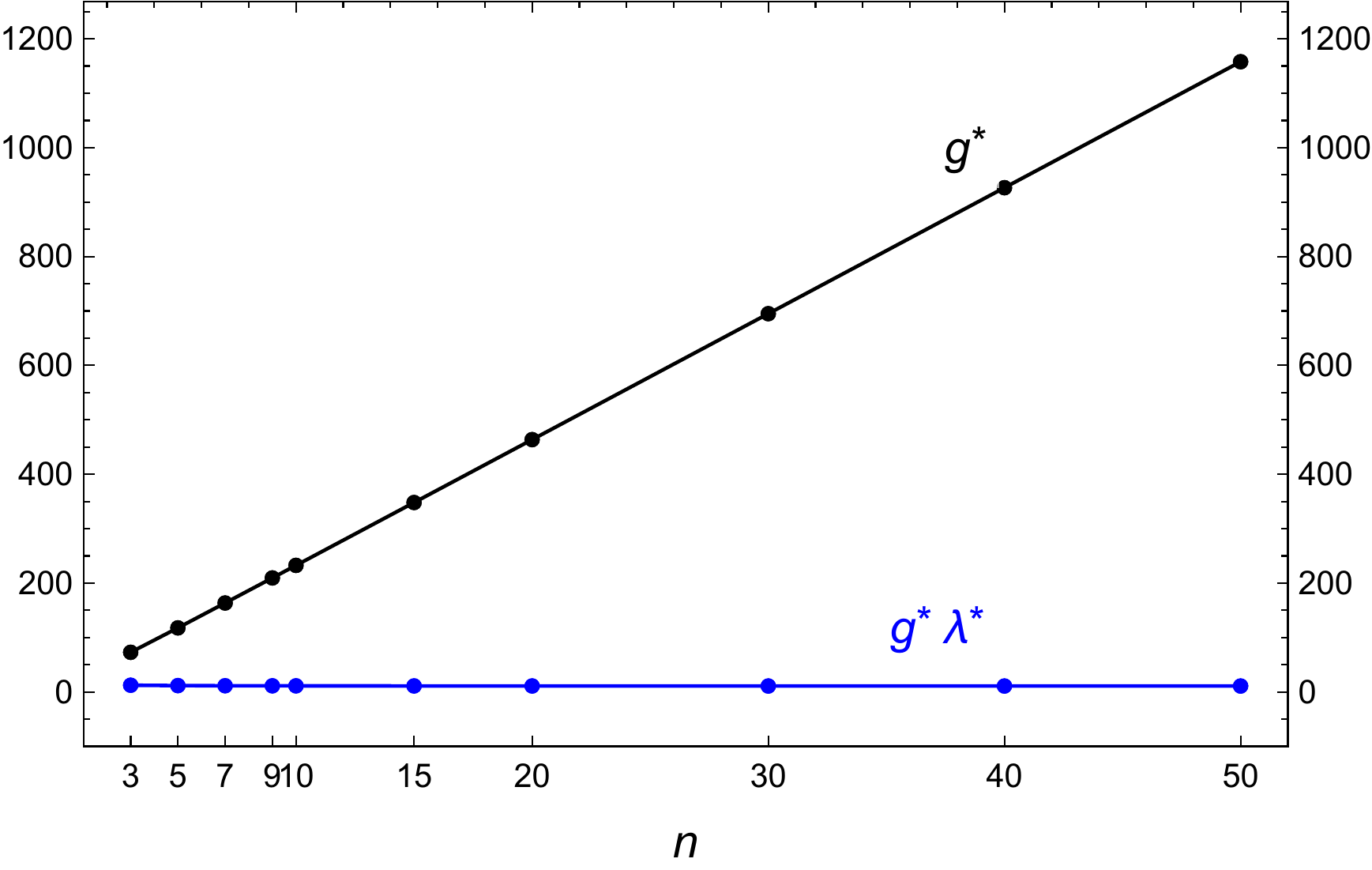}\hfill
\includegraphics[scale=0.45]{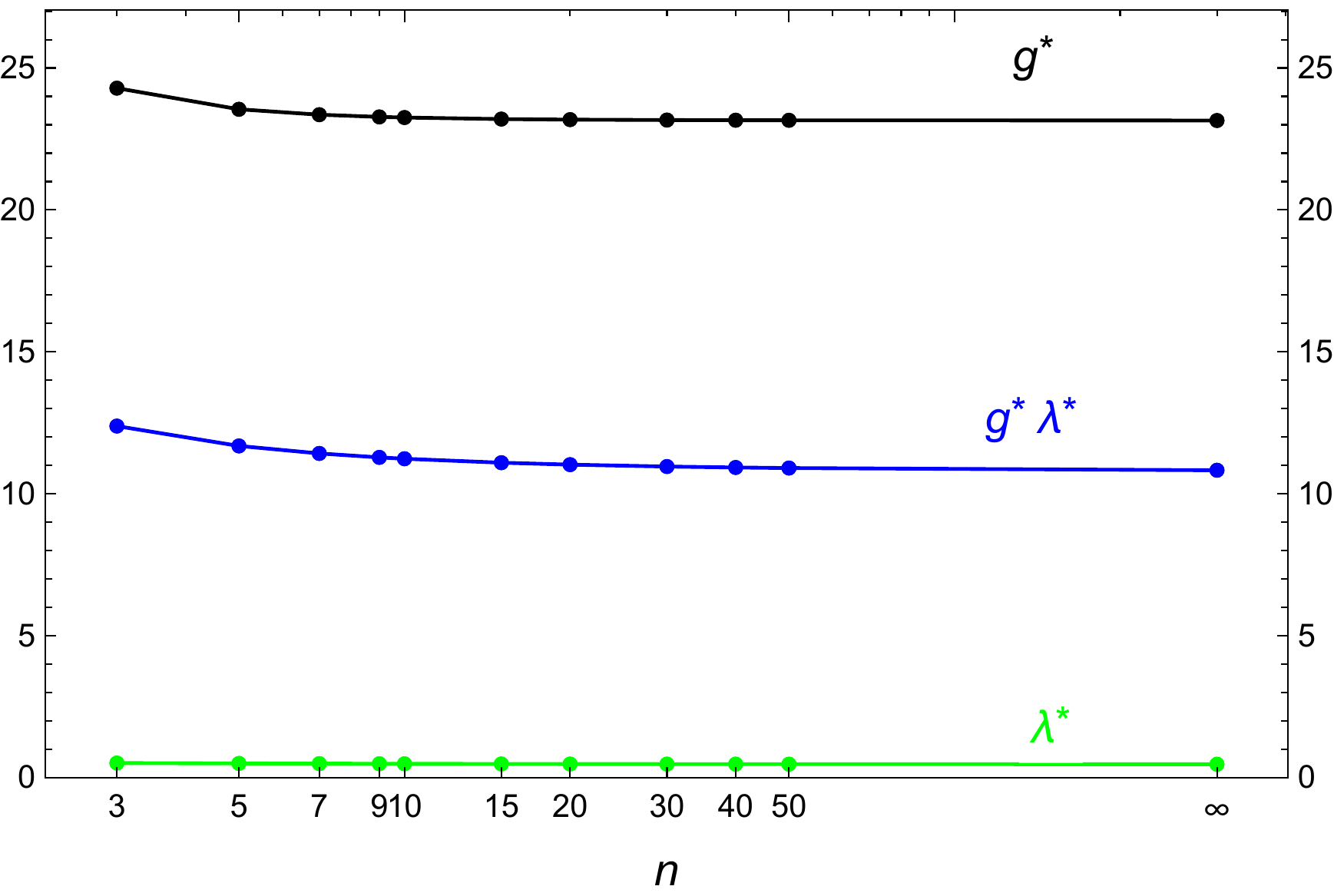}\hfill
\caption{Plot of the fixed point coordinates $g^*$, $\lambda^*$ and the product $g^*\lambda^*$ for different values of $n$. First cutoff family $\rho^{1}_k(s,n)$ implemented on the left figure, second cutoff family $\rho^{2}_k(s,n)$ implemented on the right figure.}
\label{fig:glgl_12reg}
\end{center}
\end{figure}

While the present analysis was able to show the existence of a UV attractive fixed point for the quadratically truncated theory, we also wish to compare our results with the ones of of \cite{Lauscher:2002sq}, where a similar study was conducted taking into account all the tensorial degrees of freedom of the metric and leading, similarly, to the identification of an attractive fixed point. While this main feature once again seems to be related purely to the conformal degree of the metric, we also wish to study in depth the details of the newfound fixed point through its parameters shown in Tables \ref{tablevaluesnew12PT1}-\ref{tablevaluesnew12PT3}. Before doing this, we also notice that while in the pure CREH model the UV-attractive non-Gaussian fixed point coexisted with the usual Gaussian fixed point $(g^*,\lambda^*)=(0,0)$, characterized by both an attractive and repulsive direction, in this new model the Gaussian fixed point is lost: $(g,\lambda,b)=(0,0,0)$ is a fixed point solution only for the beta functions $\partial_t g_k$ and $\partial_t \lambda_k$. This is in line with the results of \cite{Lauscher:2002sq}, where a Gaussian fixed point does not appear. Regarding the parameters in Tables \ref{tablevaluesnew12PT1}-\ref{tablevaluesnew12PT3}, we show in Figure \ref{fig:crit_2reg} the plots of the resulting critical exponents at different values of $n$. In order to perform a clear comparison with \cite{Lauscher:2002sq}, we have renamed ${\theta_3\coloneqq -\lambda_1}$ and ${-\lambda_{2,3}\coloneqq\theta'\pm i\theta''}$ to work with positive quantities. Figure \ref{fig:bandbgl_2reg} contains similar plots for the coordinate values of $b^*$ and the product $b^* g^* \lambda^*$, where the rotation $(g^*,\lambda^*,b^*)\to (-g^*,\lambda^*,-b^*)$ was performed for the same reason. As the quantities described in Figures \ref{fig:crit_2reg}-\ref{fig:bandbgl_2reg} do not change switching cutoff family, such plots are represented for only one regularizator shape $\rho_k^2(s,n)$. Figure \ref{fig:glgl_12reg} contains instead plots regarding the behaviour of the coordinates $g^*$, $\lambda^*$ and product $g^*\lambda^*$. As the coordinates $g^*$ and $\lambda^*$ depend on the regularizator shape applied, we represent the plot for each cutoff family .
\onecolumngrid
\onecolumngrid
\FloatBarrier
\FloatBarrier
\FloatBarrier
\twocolumngrid
\twocolumngrid
The curve describing the critical exponents closely resembles the behaviour found by \cite{Lauscher:2002sq} when implementing a family of exponential shape functions as regulators (see Figure 6 of \cite{Lauscher:2002sq}). On top of this, the new critical exponent $\theta_3$ is of bigger order of magnitude with respect to the other  exponents, exactly as in \cite{Lauscher:2002sq}. The curve for $b^*$  and the coordinates $g^*$, $\lambda^*$ and $g^* \lambda^*$ also share similar behaviours as in Figure 3 of \cite{Lauscher:2002sq}, further highlighting the many similarities of the two studies, in spite of the big difference in the difficulty of the problems tackled.
The only clear difference between the purely conformal theory results and the full degrees of freedom problem is shown in the right panel of Figure \ref{fig:glgl_12reg}, where the behaviour of $g^*$ decreases with $n$ instead of increasing. As we already determined that the universal quantities emerging from this analysis are the critical exponents, together with $g^*\lambda^*$, $g^*\lambda^*b^*$ and $b^*$, it does not surprise us that these are the quantities which can safely be compared between different pictures of the same model, such as this analysis and the one already performed in \cite{Lauscher:2002sq}. While strictly focusing on these parameters, the behaviours in both results share many similarities: this once again suggests that the conformal component of the metric is truly able to capture a significant portion of the UV behaviour of the model.

\section{Conclusions}
In this work we have  shown that the critical properties Reuter fixed point of a conformally reduced quadratic gravity theory like $R+R^2$ are determined by the dynamics of the conformal factor. Our approach is closer in spirit to the original paper by Reuter and Lauscher \cite{Lauscher:2002sq} it employs a linear parametrisation of the fluctuations and a compact topology for the projection of the flow equation. The role of the additional propagating degrees of freedom, if we compare our result with the full calculation presented in \cite{Lauscher:2002sq} is essentially confined to the location of the fixed point in the theory space.  On the contrary the behavior and the stability of the critical exponents against change in the regulator is qualitatively very similar to that of the full theory.  It would be very important to consider higher truncation in the CREH approximation to see if the critical properties of the theory are still determined by the conformal factor. In fact this would have far reaching consequences for the structure of the vacuum of the full theory which could be dominated by non-homogeneous field configuration which dominated the path integral as showed in \cite{Bonanno:2013dja}.  Moreover it could also shed light in a proper definition of the cutoff for the general $f(R)$ theory for which the definition of a global scaling solution strongly depend on the details of the chosen regulator \cite{Morris:2022btf}. We hope to return on these points in a following work.

\section*{Acknowledgements}
AB and MC would like to thank Benjamin Knorr for important comments and suggestions on an earlier version of the manuscript. MC would like to thank the National Institute of Astrophysics
(INAF) section of Catania and the INAF- Catania Astrophysical Observatory for warm hospitality during the preparation of the manuscript. MC acknowledges support from 
INFN.
\vspace{1.1cm}

\onecolumngrid
\clearpage
\newpage
\begin{appendices}
\section{Fixed point parameter tables}\label{APPtables}
\begin{table}[!htbp]
\centering
\begin{tabular}{|c|c|c|c|c|c|c||c|c|c|c|c|c|c|}
\hline
\multicolumn{7}{|c||}{Fixed point parameters for $\rho_k^1(s,n)$, $\beta\to 0$}&\multicolumn{7}{|c|}{Fixed point parameters for $\rho_k^2(s,n)$, $\beta\to 0$}\\
\hline
    \hline
     $\mathbf{n}$ & $\mathbf{z^*}$ & $\mathbf{g^*}$ & $\mathbf{\lambda^*}$ & $\mathbf{g^*\lambda^*}$ & $\mathbf{\theta '}$ & $\mathbf{\theta''}$ & $\mathbf{n}$ & $\mathbf{z^*}$ & $\mathbf{g^*}$ & $\mathbf{\lambda^*}$ & $\mathbf{g^*\lambda^*}$ & $\mathbf{\theta '}$ & $\mathbf{\theta''}$ \\
     \hline
     \hline
      $\mathbf{3}$ & $-0.050$ & $4.712$ & $0.375$ & $1.767$ & $-3.000$ & $4.795$ &  $\mathbf{3}$ & $-0.151$ & $1.570$ & $1.125$ & $1.767$ & $-3.000$ & $4.795$ \\
      $\mathbf{5}$ & $-0.025$ & $9.424$ & $0.250$ & $2.356$ & $-1.000$ & $3.872$  & $\mathbf{5}$ & $-0.126$ & $1.884$ & $1.250$ & $2.356$ & $-1.000$ & $3.872$         \\
      $\mathbf{7}$ & $-0.017$ & $13.482$ & $0.187$ & $2.527$ & $-0.600$ & $3.527$ & $\mathbf{7}$ & $-0.123$ & $1.926$ & $1.312$ & $2.527$ & $-0.600$ & $3.527$           \\
      $\mathbf{9}$ & $-0.013$ & $17.386$ & $0.150$ & $2.607$ & $-0.428$ & $3.353$ & $\mathbf{9}$ & $-0.123$ & $1.931$ & $1.350$  & $2.607$ & $-0.428$ & $3.353$          \\
  $\mathbf{+\infty}$ & $-$ & $-$ & $-$ & $-$ & $-$ & $-$ & $\mathbf{+\infty}$ & $-0.127$ & $1.876$ & $1.500$ & $2.815$ & $0$ & $2.828$          \\ 
      \hline
    \end{tabular}
\caption{Fixed point parameters obtained using the first cutoff $\rho_k^1(s,n)$ (on the left) and the second cutoff $\rho_k^2(s,n)$ (on the right) in the pure CREH limit case $\beta\to 0$.}\label{tablevaluesold12}
\end{table}
\FloatBarrier

\begin{table}[!htbp]
  \begin{center}
    \begin{tabular}{|c|c|c|c|c||c|c|c|c|c|}
    \hline
\multicolumn{5}{|c||}{Fixed point parameters for $\rho_k^1(s,n)$}
&\multicolumn{5}{|c|}{Fixed point parameters for $\rho_k^2(s,n)$}\\
\hline
    \hline
     $\mathbf{n}$ & $\mathbf{z^*}$ & $\mathbf{g^*}$ & $\mathbf{\lambda^*}$ & $\mathbf{b^*}$ & $\mathbf{n}$ & $\mathbf{z^*}$ & $\mathbf{g^*}$ & $\mathbf{\lambda^*}$ & $\mathbf{b^*}$ \\
     \hline
     \hline
      $\mathbf{3}$ & $3.276\times 10^{-3}$ & $-72.870$ & $0.169$ & $-3.695\times 10^{-4}$ & $\mathbf{3}$ & $9.828\times 10^{-3}$ & $-24.290$ & $0.509$ & $-3.695\times 10^{-4}$ \\
      $\mathbf{5}$ & $2.027\times 10^{-3}$ & $-117.731$ & $0.099$ & $-3.223\times 10^{-4}$ & $\mathbf{5}$ & $1.013\times 10^{-2}$ & $-23.546$ & $0.496$ & $-3.223\times 10^{-4}$ \\
      $\mathbf{7}$ & $1.460\times 10^{-3}$ & $-163.484$ & $0.069$ & $-3.061\times 10^{-4}$ & $\mathbf{7}$ & $1.022\times 10^{-2}$ & $-23.354$ & $0.488$ & $-3.061\times 10^{-4}$ \\
      $\mathbf{9}$ & $1.139\times 10^{-3}$ & $-209.495$ & $0.053$ & $-2.979\times 10^{-4}$ & $\mathbf{9}$ & $1.025\times 10^{-2}$ & $-23.277$ & $0.484$ & $-2.979\times 10^{-4}$ \\
      $\mathbf{10}$ & $1.026\times 10^{-3}$ & $-232.547$ & $0.048$ & $-2.951\times 10^{-4}$& $\mathbf{10}$ & $1.026\times 10^{-2}$ & $-23.254$ & $0.482$ & $-2.951\times 10^{-4}$ \\      
      $\mathbf{15}$ & $6.859\times 10^{-4}$ & $-348.009$ & $0.031$ & $-2.872\times 10^{-4}$ & $\mathbf{15}$ & $1.028\times 10^{-2}$ & $-23.200$ & $0.478$ & $-2.872\times 10^{-4}$ \\
      $\mathbf{20}$ & $5.149\times 10^{-4}$ & $-463.619$ & $0.023$ & $-2.835\times 10^{-4}$ & $\mathbf{20}$ & $1.029\times 10^{-2}$ & $-23.180$ & $0.475$ & $-2.835\times 10^{-4}$ \\
      $\mathbf{30}$ & $3.435\times 10^{-4}$ & $-694.983$ & $0.015$ & $-2.798\times 10^{-4}$& $\mathbf{30}$ & $1.030\times 10^{-2}$ & $-23.166$ & $0.472$ & $-2.798\times 10^{-4}$ \\                  
      $\mathbf{40}$ & $2.576\times 10^{-4}$ & $-926.417$ & $0.011$ & $-2.781\times 10^{-4}$ & $\mathbf{40}$ & $1.030\times 10^{-2}$ & $-23.160$ & $0.471$ & $-2.781\times 10^{-4}$ \\      
      $\mathbf{50}$ & $2.061\times 10^{-4}$ & $-1157.879$ & $0.009$ & $-2.770\times 10^{-4}$ & $\mathbf{50}$ & $1.030\times 10^{-2}$ & $-23.157$ & $0.470$ & $-2.770\times 10^{-4}$ \\     
      $\mathbf{+\infty}$ & $-$ & $-$ & $-$ & $-$ & $\mathbf{+\infty}$ & $1.031\times 10^{-2}$ & $-23.150$ & $0.467$ & $-2.729\times 10^{-4}$ \\       
      \hline
    \end{tabular}
     \caption{Fixed point parameters obtained using the first cutoff $\rho_k^1(s,n)$ (on the left) and the second cutoff $\rho_k^2(s,n)$ (on the right) in the quadratic theory.}\label{tablevaluesnew12PT1}
  \end{center}
\end{table}
\FloatBarrier

\begin{table}[!htbp]
  \begin{center}
    \begin{tabular}{|c|c|c|c|c||c|c|c|c|c|}
    \hline
\multicolumn{5}{|c||}{Fixed point parameters for $\rho_k^1(s,n)$}
&\multicolumn{5}{|c|}{Fixed point parameters for $\rho_k^2(s,n)$}\\
\hline
    \hline
     $\mathbf{n}$ & $\mathbf{g^*\lambda^*}$ & $\mathbf{b^*\lambda^*}$ & $\mathbf{b^*g^*}$ & $\mathbf{b^* g^* \lambda^*}$ & $\mathbf{n}$ & $\mathbf{g^*\lambda^*}$ & $\mathbf{b^*\lambda^*}$ & $\mathbf{b^*g^*}$ & $\mathbf{b^* g^* \lambda^*}$ \\
     \hline
     \hline
      $\mathbf{3}$ & $-12.382$ & $-6.279\times 10^{-5}$ & $0.026$ & $4.575\times 10^{-3}$ & $\mathbf{3}$ & $-12.382$ & $-1.883\times 10^{-4}$ & $8.975\times 10^{-3}$ & $4.575\times 10^{-3}$ \\
     $\mathbf{5}$ & $-11.682$ & $-3.198\times 10^{-5}$ & $0.037$ & $3.765\times 10^{-3}$ & $\mathbf{5}$ & $-11.682$ & $-1.599\times 10^{-4}$ & $7.589\times 10^{-3}$ & $3.765\times 10^{-3}$ \\
      $\mathbf{7}$ & $-11.418$ & $-2.138\times 10^{-5}$ & $0.050$ & $3.495\times 10^{-3}$ & $\mathbf{7}$ & $-11.418$ & $-1.496\times 10^{-4}$ & $7.149\times 10^{-3}$ & $3.495\times 10^{-3}$ \\
      $\mathbf{9}$ & $-11.279$ & $-1.604\times 10^{-5}$ & $0.062$ & $3.360\times 10^{-3}$ & $\mathbf{9}$ & $-11.279$ & $-1.443\times 10^{-4}$ & $6.934\times 10^{-3}$ & $3.360\times 10^{-3}$ \\
     $\mathbf{10}$ & $-11.231$ & $-1.425\times 10^{-5}$ & $0.068$ & $3.315\times 10^{-3}$ & $\mathbf{10}$ & $-11.231$ & $-1.425\times 10^{-4}$ & $6.864\times 10^{-3}$ & $3.315\times 10^{-3}$ \\
      $\mathbf{15}$ & $-11.092$ & $-9.156\times 10^{-6}$ & $0.099$ & $3.186\times 10^{-3}$ & $\mathbf{15}$ & $-11.092$ & $-1.373\times 10^{-4}$ & $6.665\times 10^{-3}$ & $3.186\times 10^{-3}$ \\
     $\mathbf{20}$ & $-11.024$ & $-6.741\times 10^{-6}$ & $0.131$ & $3.125\times 10^{-3}$ & $\mathbf{20}$ & $-11.024$ & $-1.348\times 10^{-4}$ & $6.572\times 10^{-3}$ & $3.125\times 10^{-3}$ \\
     $\mathbf{30}$ & $-10.957$  & $-4.412\times 10^{-6}$ & $0.194$ & $3.066\times 10^{-3}$ & $\mathbf{30}$ & $-10.957$ & $-1.323\times 10^{-4}$ & $6.483\times 10^{-3}$ & $3.066\times 10^{-3}$ \\
     $\mathbf{40}$ & $-10.924$ & $-3.279\times 10^{-6}$ & $0.257$ & $3.038\times 10^{-3}$ & $\mathbf{40}$ & $-10.924$ & $-1.311\times 10^{-4}$ & $6.441\times 10^{-3}$ & $3.038\times 10^{-3}$ \\
     $\mathbf{50}$ & $-10.904$ & $-2.609\times 10^{-6}$ & $0.320$ & $3.021\times 10^{-3}$ & $\mathbf{50}$ & $-10.904$ & $-1.304\times 10^{-4}$ & $6.415\times 10^{-3}$ & $3.021\times 10^{-3}$ \\
     $\mathbf{+\infty}$ & $-$ & $-$ & $-$ & $-$ & $\mathbf{+\infty}$ & $-10.826$ & $-1.276\times 10^{-4}$ & $6.318\times 10^{-3}$ & $2.954\times 10^{-3}$ \\
      \hline
    \end{tabular}
     \caption{Fixed point parameters obtained using the first cutoff $\rho_k^1(s,n)$ (on the left) and the second cutoff $\rho_k^2(s,n)$ (on the right) in the quadratic theory.}\label{tablevaluesnew12PT2}
  \end{center}
\end{table}
\FloatBarrier

\begin{table}[!htbp]
  \begin{center}
    \begin{tabular}{|c|c|c|c||c|c|c|c|c|}
    \hline
\multicolumn{4}{|c||}{Fixed point parameters for $\rho_k^1(s,n)$}
&\multicolumn{4}{|c|}{Fixed point parameters for $\rho_k^2(s,n)$}\\
\hline
    \hline
     $\mathbf{n}$ & $\mathbf{\lambda_1}$ &  $\mathbf{\theta'}$ &  $\mathbf{\theta''}$ & $\mathbf{n}$ & $\mathbf{\lambda_1}$ &  $\mathbf{\theta'}$ &  $\mathbf{\theta''}$ \\
     \hline
     \hline
      $\mathbf{3}$ & $-3.035$ & $-0.584$ & $1.488$ & $\mathbf{3}$ & $-3.035$ & $-0.584$ & $1.488$  \\
     $\mathbf{5}$ & $-2.546$ & $-0.586$ & $1.494$ & $\mathbf{5}$ & $-2.546$ & $-0.586$ & $1.494$ \\
      $\mathbf{7}$ &   $-2.382$ & $-0.588$ & $1.500$ & $\mathbf{7}$ & $-2.382$ & $-0.588$ & $1.500$ \\
      $\mathbf{9}$ & $-2.300$ & $-0.589$ & $1.504$ & $\mathbf{9}$ & $-2.300$ & $-0.589$ & $1.504$ \\
     $\mathbf{10}$ & $-2.272$ & $-0.589$ & $1.505$ & $\mathbf{10}$ & $-2.272$ & $-0.589$ & $1.505$\\
      $\mathbf{15}$ & $-2.193$ & $-0.590$ & $1.510$ & $\mathbf{15}$ & $-2.193$ & $-0.590$ & $1.510$ \\
     $\mathbf{20}$ & $-2.156$ & $-0.591$ & $1.512$ & $\mathbf{20}$  & $-2.156$ & $-0.591$ & $1.512$ \\
     $\mathbf{30}$ & $-2.120$ & $-0.592$ & $1.515$ & $\mathbf{30}$ & $-2.120$ & $-0.592$ & $1.515$ \\
     $\mathbf{40}$ & $-2.103$ & $-0.592$ & $1.516$ & $\mathbf{40}$ & $-2.103$ & $-0.592$ & $1.516$ \\
     $\mathbf{50}$ & $-2.092$ & $-0.592$ & $1.517$ & $\mathbf{50}$ & $-2.092$ & $-0.592$ & $1.517$ \\
     $\mathbf{+\infty}$ & $-$ & $-$ & $-$ & $\mathbf{+\infty}$ & $-2.503$ & $-0.709$ & $1.966$ \\
      \hline
    \end{tabular}
     \caption{Fixed point parameters obtained using the first cutoff $\rho_k^1(s,n)$ (on the left) and the second cutoff $\rho_k^2(s,n)$ (on the right) in the quadratic theory.}\label{tablevaluesnew12PT3}
  \end{center}
\end{table}
\FloatBarrier
\section{Explicit expressions for beta functions}\label{APPbetafuncs}
In this section, we explicitly report some explicit expressions for the beta functions obtained for dimension $d=4$ using the second cutoff $\rho_k^2(s,n)$ at different values for the smoothness parameter $n$ (finite $n=\left\lbrace3,5,7,9\right\rbrace$ and the limit $n\to\infty$). We do not show the remaining $n=\left\lbrace10,15,20,30,40,50\right\rbrace$ as their expressions get extremely cumbersome. Each choice of $n$ leads to couples of beta functions depending on the sign of the coupling $\beta$. This section contains beta functions only related to the sign $\beta<0$, which are the ones leading to significant physical results as shown in Section \ref{Results}. Beta functions for $\beta>0$ are not shown here due to excessive length. To avoid further prolixity, we do not include the explicit expressions for the beta functions pertaining the choice of the first cutoff $\rho_k^1(s,n)$, as they share similar structure to their second cutoff counterparts.
\subsubsection{Beta functions for dimensionless couplings at $d=4$, with second cutoff $\rho_k^2(s,n)$, smoothness parameter $n=3$}
\small
\begin{align}
&\partial_t z=-\frac{1}{16 \pi ^2 (3-2 \lambda )^2 (288 b (2 \lambda -3)+z)^4}\left(220150628352 \pi ^2 b^4 (3-2 \lambda )^6 z+z^4 \left(32 \pi ^2 (3-2 \lambda )^2 z+27\right)+\right.\nonumber\\
&+144 b (2 \lambda -3) z^3 \left(256 \pi ^2 (3-2 \lambda )^2 z+351\right)+\nonumber\\
&+5971968 b^3 (2 \lambda -3)^3 \sqrt{z} \left(-81  \pi  \sqrt{288 b (3-2 \lambda )-z}+512 \pi ^2 (3-2 \lambda )^2 z^{3/2}-1134 \sqrt{z}\right)+\nonumber\\
&+62208 b^2 (3-2 \lambda )^2 z^{3/2} \left(243  \pi  \sqrt{288 b (3-2 \lambda )-z}+256 \pi ^2 (3-2 \lambda )^2 z^{3/2}-180 \sqrt{z}\right)+\nonumber\\
&\left.-30233088 b^2 (3-2 \lambda )^2 \sqrt{z} (32 b (2 \lambda -3)-z) \sqrt{288 b (2 \lambda -3)+z} \tanh ^{-1}\left(\frac{\sqrt{z}}{\sqrt{288 b (2 \lambda -3)+z}}\right)\right)\nonumber\\
&\partial_t \lambda=\frac{1}{16 \pi ^2 (3-2 \lambda )^2 \sqrt{z} (288 b (2 \lambda -3)+z)^4}\left(z^{7/2} \left(\lambda  \left(189-32 \pi ^2 (3-2 \lambda )^2 z\right)-243\right)-220150628352 \pi ^2 b^4 (3-2 \lambda )^6 \lambda  \sqrt{z}+\right.\nonumber\\
&-5971968 b^3 (2 \lambda -3)^3 \left(81  \pi  \lambda  \sqrt{288 b (3-2 \lambda)-z}+512 \pi ^2 \lambda  (3-2 \lambda )^2 z^{3/2}+486 (5 \lambda -4) \sqrt{z}\right)+\nonumber\\
&+144 b (2 \lambda -3) z^2 \left(243  \pi  (2 \lambda -3) \sqrt{288 b (3-2 \lambda)-z}-256 \pi ^2 \lambda  (3-2 \lambda )^2 z^{3/2}+351 \lambda  \sqrt{z}\right)+\nonumber\\
&-62208 b^2 (3-2 \lambda )^2 z \left(-81  \pi  (7 \lambda -6) \sqrt{288 b (3-2 \lambda )-z}+256 \pi ^2 \lambda  (3-2 \lambda )^2 z^{3/2}+36 (23 \lambda -27)
   \sqrt{z}\right)+\nonumber\\
&\left.-69984 b (3-2 \lambda )^2 \sqrt{288 b (2 \lambda -3)+z} \left(13824 b^2 \lambda  (2 \lambda -3)+144 b (6-7 \lambda ) z-z^2\right) \tanh
   ^{-1}\left(\frac{\sqrt{z}}{\sqrt{288 b (2 \lambda -3)+z}}\right)\right)\nonumber\\
&\partial_t b=-\frac{3}{256 \pi ^2 (288 b (2 \lambda -3)+z)^5}\left(3962711310336 b^5 (3-2 \lambda )^2+\frac{2021760 b^2 z^3}{2 \lambda -3}+\frac{1944 b z^4}{(3-2 \lambda )^2}+\frac{z^5}{(2 \lambda -3)^3}+\right.\nonumber\\
&-4478976 b^3 z^{3/2} \left(96 \sqrt{z}-115  \pi  \sqrt{288 b (3-2 \lambda )-z}\right)-429981696  b^4 (2 \lambda -3) \left(75 \pi  \sqrt{z} \sqrt{288 b (3-2 \lambda
   )-z}+554  z\right)+\nonumber\\
&\left.+44789760 b^3 \sqrt{z} (23 z-1440 b (2 \lambda -3)) \sqrt{288 b (2 \lambda -3)+z} \tanh ^{-1}\left(\frac{\sqrt{z}}{\sqrt{288 b (2 \lambda -3)+z}}\right)\right)\nonumber      
\end{align}
\normalsize
\subsubsection{Beta functions for dimensionless couplings at $d=4$, with second cutoff $\rho_k^2(s,n)$, smoothness parameter $n=5$}
\small
\begin{align}
&\partial_t z=-\frac{1}{32 \pi ^2 (5-2 \lambda )^4 (288 b (2 \lambda -5)+z)^6}\left( z^6 \left(64 \pi ^2 (5-2 \lambda )^4 z+3125\right)+ \right.\nonumber\\
&+51840 b^2 (5-2 \lambda )^2 z^4 \left(1536 \pi ^2 (5-2 \lambda )^4 z+105625\right)+36520347436056576 \pi ^2 b^6 (5-2 \lambda )^{10} z+\nonumber\\
&+7464960 b^3 (2 \lambda -5)^3 z^3 \left(4096 \pi ^2 (5-2 \lambda )^4 z+433125\right)+48 b (2 \lambda -5) z^5 \left(2304 \pi ^2 (5-2 \lambda )^4 z+128125\right)+\nonumber\\
&+61917364224 b^5 (2 \lambda -5)^5 \sqrt{z} \left(-109375  \pi  \sqrt{288 b (5-2 \lambda)-z}+12288 \pi ^2 (5-2 \lambda )^4 z^{3/2}-2181250 \sqrt{z}\right)+\nonumber\\
&+358318080 b^4 (5-2 \lambda )^4 z^{3/2} \left(1115625  \pi  \sqrt{288 b (5-2 \lambda )-z}+18432 \pi ^2 (5-2 \lambda )^4 z^{3/2}+372500 \sqrt{z}\right)+\nonumber\\
&\left.-47029248000000 b^4 (5-2 \lambda )^4 \sqrt{z} (288 b (2 \lambda -5)-17 z) \sqrt{288 b (2 \lambda -5)+z} \tanh ^{-1}\left(\frac{\sqrt{z}}{\sqrt{288 b (2 \lambda
   -5)+z}}\right)\right)\nonumber\\
&\partial_t \lambda=\frac{1}{32 \pi ^2 (5-2 \lambda )^4 \sqrt{z} (288 b (2 \lambda -5)+z)^6}\left(-36520347436056576 \pi ^2 b^6 (5-2 \lambda )^{10} \lambda  \sqrt{z}+ \right.\nonumber\\
&   +z^{11/2} \left(\lambda  \left(9375-64 \pi ^2 (5-2 \lambda )^4 z\right)-15625\right)-24 b (2 \lambda -5) z^{9/2} \left(4 \lambda  \left(1152 \pi ^2 (5-2 \lambda )^4
   z-190625\right)+1265625\right)+\nonumber\\
&-51840 b^2 (5-2 \lambda )^2 z^{7/2} \left(\lambda  \left(1536 \pi ^2 (5-2 \lambda )^4 z-319375\right)+534375\right)+\nonumber\\
&-61917364224 b^5 (2 \lambda -5)^5 \left(109375  \pi  \lambda  \sqrt{288 b (5-2 \lambda )-z}+12288 \pi ^2 \lambda  (5-2 \lambda )^4 z^{3/2}+6250 (541 \lambda -480)
   \sqrt{z}\right)+\nonumber\\
&-7464960 b^3 (2 \lambda -5)^3 z^2 \left(-65625  \pi  (2 \lambda -5) \sqrt{288 b (5-2 \lambda )-z}+4096 \pi ^2 \lambda  (5-2 \lambda )^4 z^{3/2}+625 (1640-1349 \lambda
   ) \sqrt{z}\right)+\nonumber\\
&+358318080 b^4 (5-2 \lambda )^4 z \left(65625  \pi  (29 \lambda -30) \sqrt{288 b (5-2 \lambda )-z}-18432 \pi ^2 \lambda  (5-2 \lambda )^4 z^{3/2}+12500 (19 \lambda
   +27) \sqrt{z}\right)+\nonumber\\
& -979776000000 b^3 (5-2 \lambda )^4 \sqrt{288 b (2 \lambda -5)+z}\times\nonumber\\ 
&\left.\times\left(13824 b^2 \lambda  (2 \lambda -5)+48 b (30-29 \lambda ) z-z^2\right) \tanh
   ^{-1}\left(\frac{\sqrt{z}}{\sqrt{288 b (2 \lambda -5)+z}}\right)\right)\nonumber\\
&\partial_t b=-\frac{3125}{2304 \pi ^2 (288 b (2 \lambda -5)+z)^7}\left(328683126924509184 b^7 (5-2 \lambda )^2+\frac{1474702848 b^3 z^4}{(5-2 \lambda )^2}+\frac{2268 b z^6}{(5-2 \lambda )^4}+\right.\nonumber\\
&+\frac{2327616 b^2 z^5}{(2 \lambda -5)^3}+\frac{731775098880 b^4 z^3}{2 \lambda -5}+7739670528 b^5 z^{3/2} \left(32 \sqrt{z}+12285  \pi  \sqrt{288 b (5-2 \lambda
   )-z}\right)+\frac{z^7}{(2 \lambda -5)^5}+\nonumber\\
&-2229025112064  b^6 (2 \lambda -5) \left(1575 \pi  \sqrt{z} \sqrt{288 b (5-2 \lambda )-z}+16306  z\right)+\nonumber\\
&+\left.14627977297920 b^5 \sqrt{z} (13 z-480 b (2 \lambda -5)) \sqrt{288 b (2 \lambda -5)+z} \tanh ^{-1}\left(\frac{\sqrt{z}}{\sqrt{288 b (2 \lambda -5)+z}}\right)\right)\nonumber             
\end{align}
\normalsize
\subsubsection{Beta functions for dimensionless couplings at $d=4$, with second cutoff $\rho_k^2(s,n)$, smoothness parameter $n=7$}
\small
\begin{align}
&\partial_t z=-\frac{1}{240 \pi ^2 (7-2 \lambda )^6 (288 b (2 \lambda -7)+z)^8}\left(5 z^8 \left(96 \pi ^2 (7-2 \lambda )^6 z+823543\right)+\right.\nonumber\\
&+36288 b^2 (7-2 \lambda )^2 z^6 \left(30720 \pi ^2 (7-2 \lambda )^6 z+304357963\right)+\nonumber\\
&+376233984 b^4 (7-2 \lambda )^4 z^4 \left(614400 \pi ^2 (7-2 \lambda )^6 z+8405197507\right)+\nonumber\\
&+22718577733022074798080 \pi ^2 b^8 (7-2 \lambda )^{14} z+720 b (2 \lambda -7) z^7 \left(1536 \pi ^2 (7-2 \lambda )^6 z+14000231\right)+\nonumber\\
&+15676416 b^3 (2 \lambda -7)^3 z^5 \left(40960 \pi ^2 (7-2 \lambda )^6 z+458713451\right)+\nonumber\\
&+54177693696 b^5 (2 \lambda -7)^5 z^3 \left(983040 \pi ^2 (7-2 \lambda )^6 z+20161156183\right)+\nonumber\\
&+4814694242058240 b^7 (2 \lambda -7)^7 \sqrt{z} \left(190238433  \pi  \sqrt{288 b (7-2 \lambda )-z}+131072 \pi ^2 (7-2 \lambda )^6 z^{3/2}-4679371326 \sqrt{z}\right)+\nonumber\\
&+39007939461120 b^6 (7-2 \lambda )^6 z^{3/2} \left(-2038268925  \pi  \sqrt{288 b (7-2 \lambda )-z}+196608 \pi ^2 (7-2 \lambda )^6 z^{3/2}+2242389940 \sqrt{z}\right)+\nonumber\\
&\left. -6360693666550571335680 b^6 (7-2 \lambda )^6 \sqrt{z} (288 b (2 \lambda -7)-25 z) \sqrt{288 b (2 \lambda -7)+z} \tanh ^{-1}\left(\frac{\sqrt{z}}{\sqrt{288 b (2 \lambda
   -7)+z}}\right)\right)\nonumber\\
&\partial_t \lambda=\frac{1}{240 \pi ^2 (7-2 \lambda )^6 \sqrt{z} (288 b (2 \lambda -7)+z)^8}\left(-22718577733022074798080 \pi ^2 b^8 (7-2 \lambda )^{14} \lambda  \sqrt{z}+\right.\nonumber\\
&+z^{15/2} \left(\lambda  \left(9058973-480 \pi ^2 (7-2 \lambda )^6 z\right)-17294403\right)+\nonumber\\
&-7838208 b^3 (2 \lambda -7)^3 z^{9/2} \left(20 \lambda  \left(4096 \pi ^2 (7-2 \lambda )^6 z-98236915\right)+3665589893\right)+\nonumber\\
&-36 b (2 \lambda -7) z^{13/2} \left(2 \lambda  \left(15360 \pi ^2 (7-2 \lambda )^6 z-305534453\right)+1158725001\right)+\nonumber\\
&-36288 b^2 (7-2 \lambda )^2 z^{11/2} \left(\lambda  \left(30720 \pi ^2 (7-2 \lambda )^6 z-658010857\right)+1237785129\right)+\nonumber\\
&-376233984 b^4 (7-2 \lambda )^4 z^{7/2} \left(\lambda  \left(614400 \pi ^2 (7-2 \lambda )^6 z-18035709349\right)+33706791447\right)+\nonumber\\
&+4814694242058240 b^7 (7-2 \lambda )^7 \left(131072 \pi ^2 \lambda  (7-2 \lambda )^6 z^{3/2}+1647086 (3865 \lambda -3584) \sqrt{z}\right)+\nonumber\\
&-915939887983282272337920  \pi  b^7 (7-2 \lambda )^7 \lambda  \sqrt{288 b (7-2 \lambda )-z}+\nonumber\\
&-54177693696 b^5 (2 \lambda -7)^5 z^2 \left(983040 \pi ^2 (7-2 \lambda )^6 \lambda  z^{3/2}-117649 (300247 \lambda -451080) \sqrt{z}\right)+\nonumber\\
&-66257225693235118080  \pi  b^5 (2 \lambda -7)^6 z^2 \sqrt{288 b (7-2 \lambda )-z}+\nonumber\\
&-39007939461120 b^6 (7-2 \lambda )^6 z \left(196608 \pi ^2 (7-2 \lambda )^6 \lambda  z^{3/2}-470596 (6751 \lambda -6951) \sqrt{z}\right)+\nonumber\\
&-3180346833275285667840  \pi  b^6 (7-2 \lambda )^6 (37 \lambda -42) z \sqrt{288 b (7-2 \lambda )-z}+\nonumber\\
&-132514451386470236160 b^5 (7-2 \lambda )^6 \sqrt{288 b (2 \lambda -7)+z}\times\nonumber\\
&\left.\times \left(13824 b^2 \lambda  (2 \lambda -7)+48 b (42-37 \lambda ) z-z^2\right) \tanh ^{-1}\left(\frac{\sqrt{z}}{\sqrt{288 b (2 \lambda -7)+z}}\right)\right)\nonumber\\
&\partial_t b=-\frac{823543}{2304 \pi ^2 (288 b (2 \lambda -7)+z)^9}\left(\frac{2597432832 b^3 z^6}{(7-2 \lambda )^4}+\frac{496573391241216 b^5 z^4}{(7-2 \lambda )^2}+\frac{2760 b z^8}{(7-2 \lambda )^6}+\right.\nonumber\\
&+27262293279626489757696 b^9 (7-2 \lambda )^2+\frac{1322964099072 b^4 z^5}{(2 \lambda -7)^3}+\frac{3452544 b^2 z^7}{(2 \lambda -7)^5}+\frac{z^9}{(2 \lambda -7)^7}+\nonumber\\
&+\frac{159936452584538112 b^6 z^3}{2 \lambda -7}+69331597085638656  b^8 (2 \lambda -7) \sqrt{z} \left(5005 \pi  \sqrt{288 b (7-2 \lambda )-z}-63718  \sqrt{z}\right)+\nonumber\\
&+26748301344768 b^7 \left(312896 z^2-495495  \pi  z^{3/2} \sqrt{288 b (7-2 \lambda )-z}\right)+\nonumber\\
&\left.-2409754468150149120 b^7 \sqrt{z} (288 b (2 \lambda -7)-11 z) \sqrt{288 b (2 \lambda -7)+z} \tanh ^{-1}\left(\frac{\sqrt{z}}{\sqrt{288 b (2 \lambda -7)+z}}\right)\right)\nonumber
\end{align}
\normalsize
\subsubsection{Beta functions for dimensionless couplings at $d=4$, with second cutoff $\rho_k^2(s,n)$, smoothness parameter $n=9$}
\small
\begin{align}
&\partial_t z=-\frac{1}{448 \pi ^2 (9-2 \lambda )^8 (288 b (2 \lambda -9)+z)^{10}}\left(7 z^{10} \left(128 \pi ^2 (9-2 \lambda )^8 z+387420489\right)+\right.\nonumber\\
&+93312 b^2 (9-2 \lambda )^2 z^8 \left(35840 \pi ^2 (9-2 \lambda )^8 z+117474501609\right)+\nonumber\\
&+1128701952 b^4 (9-2 \lambda )^4 z^6 \left(1146880 \pi ^2 (9-2 \lambda )^8 z+4320642433491\right)+\nonumber\\
&+8358844170240 b^6 (9-2 \lambda )^6 z^4 \left(12845056 \pi ^2 (9-2 \lambda )^8 z+66078826024329\right)+\nonumber\\
&+3517490128110528214496968704 \pi ^2 b^{10} (9-2 \lambda )^{18} z+720 b (2 \lambda -9) z^9 \left(3584 \pi ^2 (9-2 \lambda )^8 z+11235194181\right)+\nonumber\\
&+4478976 b^3 (2 \lambda -9)^3 z^7 \left(573440 \pi ^2 (9-2 \lambda )^8 z+1993020135579\right)+\nonumber\\
&+69657034752 b^5 (2 \lambda -9)^5 z^5 \left(6422528 \pi ^2 (9-2 \lambda )^8 z+27236736544725\right)+\nonumber\\
&+8425714923601920 b^7 (2 \lambda -9)^7 z^3 \left(2097152 \pi ^2 (9-2 \lambda )^8 z+15922594677411\right)+\nonumber\\
&+23295416620774588416 b^9 (2 \lambda -9)^9 \sqrt{z} \left(5242880 \pi ^2 (9-2 \lambda )^8 z^{3/2}-71183865807882 \sqrt{z}\right)+\nonumber\\
&+58076658131000771709664430653440  \pi  b^9 (2 \lambda -9)^9 \sqrt{z} \sqrt{288 b (9-2 \lambda )-z}+\nonumber\\
&+727981769399205888 b^8 (9-2 \lambda )^8 z^{3/2} \left(2621440 \pi ^2 (9-2 \lambda )^8 z^{3/2}+15755616446652 \sqrt{z}\right)+\nonumber\\
&-6654617077510505091732382679040  \pi  b^8 (9-2 \lambda )^8 z^{3/2} \sqrt{288 b (9-2 \lambda )-z}+\nonumber\\
&-1209930377729182743951342305280 b^8 (9-2 \lambda )^8 \sqrt{z} (96 b (2 \lambda -9)-11 z)\times\nonumber\\
&\times \left. \sqrt{288 b (2 \lambda -9)+z} \tanh ^{-1}\left(\frac{\sqrt{z}}{\sqrt{288 b (2 \lambda -9)+z}}\right)\right)\nonumber\\
&\partial_t \lambda=\frac{1}{448 \pi ^2 (9-2 \lambda )^8 \sqrt{z} (288 b (2 \lambda -9)+z)^{10}}\left(-3517490128110528214496968704 \pi ^2 b^{10} (9-2 \lambda )^{18} \lambda  \sqrt{z}+\right.\nonumber\\
&+z^{19/2} \left(\lambda  \left(5036466357-896 \pi ^2 (9-2 \lambda )^8 z\right)-10460353203\right)+\nonumber\\
&-72 b (2 \lambda -9) z^{17/2} \left(8 \lambda  \left(4480 \pi ^2 (9-2 \lambda )^8 z-25957172763\right)+428874481323\right)+\nonumber\\
&-93312 b^2 (9-2 \lambda )^2 z^{15/2} \left(5 \lambda  \left(7168 \pi ^2 (9-2 \lambda )^8 z-43175861163\right)+442821618927\right)+\nonumber\\
&-1119744 b^3 (2 \lambda -9)^3 z^{13/2} \left(2 \lambda  \left(1146880 \pi ^2 (9-2 \lambda )^8 z-7274077961301\right)+29592339211287\right)+\nonumber\\
&-161243136 b^4 (9-2 \lambda )^4 z^{11/2} \left(5 \lambda  \left(1605632 \pi ^2 (9-2 \lambda )^8 z-10946049356043\right)+110185873856001\right)+\nonumber\\
&-34828517376 b^5 (2 \lambda -9)^5 z^{9/2} \left(4 \lambda  \left(3211264 \pi ^2 (9-2 \lambda )^8 z-24412871647125\right)+194301060745725\right)+\nonumber\\
&-8358844170240 b^6 (9-2 \lambda )^6 z^{7/2} \left(\lambda  \left(12845056 \pi ^2 (9-2 \lambda )^8 z-118001694801087\right)+233652909495411\right)+\nonumber\\
&+23295416620774588416 b^9 (9-2 \lambda )^9 \left(5242880 \pi ^2 \lambda  (9-2 \lambda )^8 z^{3/2}+2324522934 (38815 \lambda -36864) \sqrt{z}\right)+\nonumber\\
&-58076658131000771709664430653440  \pi  b^9 (9-2 \lambda )^9 \lambda  \sqrt{288 b (9-2 \lambda )-z}+\nonumber\\
&+1685142984720384 b^7 (9-2 \lambda )^7 z^2 \left(10485760 \pi ^2 (9-2 \lambda )^8 \lambda  z^{3/2}-1162261467 (108941 \lambda -181992) \sqrt{z}\right)+\nonumber\\
&+4201147144892995638719938560  \pi  b^7 (9-2 \lambda )^7 (2 \lambda -9) z^2 \sqrt{288 b (9-2 \lambda )-z}+\nonumber\\
&-727981769399205888 b^8 (9-2 \lambda )^8 z \left(2621440 \pi ^2 (9-2 \lambda )^8 \lambda  z^{3/2}-1549681956 (14069 \lambda -17559) \sqrt{z}\right)+\nonumber\\
&-1814895566593774115927013457920  \pi  b^8 (9-2 \lambda )^8 (5 \lambda -6) z \sqrt{288 b (9-2 \lambda )-z}+\nonumber\\
&-8402294289785991277439877120 b^7 (9-2 \lambda )^8 \sqrt{288 b (2 \lambda -9)+z}\times\nonumber\\
&\left.\times \left(13824 b^2 \lambda  (2 \lambda -9)-432 b (5 \lambda -6) z-z^2\right) \tanh ^{-1}\left(\frac{\sqrt{z}}{\sqrt{288 b (2 \lambda -9)+z}}\right)\right)\nonumber\\
&\partial_t b=-\frac{43046721}{1792 \pi ^2 (288 b (2 \lambda -9)+z)^{11}}\left(\frac{32005082880 b^3 z^8}{(9-2 \lambda )^6}+\frac{8988606649221120 b^5 z^6}{(9-2 \lambda )^4}+\frac{807924802135145840640 b^7 z^4}{(9-2 \lambda )^2}+\right.\nonumber\\
&+15828705576497376965236359168 b^{11} (9-2 \lambda )^2+\frac{19974450327552 b^4 z^7}{(2 \lambda -9)^5}+\frac{34844256 b^2 z^9}{(2 \lambda -9)^7}+\frac{7 z^{11}}{(2
   \lambda -9)^9}+\nonumber\\
&+\frac{191751763025568454410240 b^8 z^3}{2 \lambda -9}+\frac{3037835650762801152 b^6 z^5}{(2 \lambda -9)^3}+\frac{23058 b
   z^{10}}{(9-2 \lambda )^8}+\nonumber\\
&+2096587495869712957440  b^{10} (2 \lambda -9) \sqrt{z} \left(109395 \pi  \sqrt{288 b (9-2 \lambda )-z}-1616218 \sqrt{z}\right)\nonumber\\
&+1455963538798411776 b^9 \left(8947808 z^2-7767045  \pi  z^{3/2} \sqrt{288 b (9-2 \lambda )-z}\right)+\nonumber\\
&-\left.318550262653704512471040 b^9 \sqrt{z} \sqrt{288 b (2 \lambda -9)+z} (1440 b (2 \lambda -9)-71 z) \tanh ^{-1}\left(\frac{\sqrt{z}}{\sqrt{288 b (2 \lambda
   -9)+z}}\right)\right)\nonumber
\end{align}
\normalsize
\subsubsection{Beta functions for dimensionless couplings at $d=4$, with second cutoff $\rho_k^2(s,n)$, smoothness parameter $n\to\infty$}
\small
\begin{align}
&\partial_t z=\frac{\sqrt{2 \pi } \sqrt{z} (72 b+z) \text{erfc}\left(\frac{\sqrt{z}}{12 \sqrt{2} \sqrt{b}}\right) e^{\frac{z}{288 b}+2 \lambda }-24 \sqrt{b} z \left(1152 \pi ^2
   b+e^{2 \lambda }\right)}{13824 \pi ^2 b^{3/2}}\nonumber\\
   \nonumber\\
&\partial_t \lambda=\frac{-\sqrt{2 \pi } \text{erfc}\left(\frac{\sqrt{z}}{12 \sqrt{2} \sqrt{b}}\right) e^{\frac{z}{288 b}+2 \lambda } (2 \lambda  (72 b+z)-3 z)-24 \sqrt{b} \sqrt{z}
   \left(2304 \pi ^2 b \lambda +e^{2 \lambda } (3-2 \lambda )\right)}{27648 \pi ^2 b^{3/2} \sqrt{z}}\nonumber\\
   \nonumber\\
&\partial_t b=\frac{e^{2 \lambda } \left(24 \sqrt{b} (72 b+z)-\sqrt{2 \pi } \sqrt{z} e^{\frac{z}{288 b}} (180 b+z) \text{erfc}\left(\frac{\sqrt{z}}{12 \sqrt{2}
   \sqrt{b}}\right)\right)}{995328 \pi ^2 b^{3/2}}\nonumber
\end{align}
\normalsize

\end{appendices}
\bibliography{ch}
\end{document}